\documentclass[aps,prl,reprint,superscriptaddress,longbibliography]{revtex4-1}
\usepackage[english]{babel}
\usepackage{siunitx}  
\usepackage{graphicx} 
\usepackage{miller}   
\usepackage[percent]{overpic} 
\usepackage{pifont}
\usepackage{amssymb} 
\usepackage{csquotes} 
\usepackage{xcolor}  
\usepackage[utf8]{inputenc}
\DeclareSIUnit\rydberg{Ry}

\begin{document}

\title{Effect of strain and many-body corrections on the band inversions and topology of bismuth}

\author{Christian \surname{König}}
\affiliation{Tyndall National Institute, University College Cork, Lee Maltings, Cork T12 R5CP, Ireland}
\author{James C. \surname{Greer}}
\affiliation{Nottingham Ningbo New Material Institute and Department of Electrical and Electronic Engineering, University of Nottingham Ningbo China, 199 Taikang East Road, Ningbo, 315100, China}
\author{Stephen \surname{Fahy}}
\affiliation{Tyndall National Institute, University College Cork, Lee Maltings, Cork T12 R5CP, Ireland}
\affiliation{Department of Physics, University College Cork, College Road, Cork T12 K8AF, Ireland}

\date{\today}

\begin{abstract}
The electronic band structure of Bi is calculated using state of the
art electronic structure methods, including density functional theory
and G$_0$W$_0$ quasiparticle approximations. The delicate ordering of
states at the L point of the Brillouin zone, which determines the
topological character of the electronic bands, is investigated in
detail. The effect on the bands of strain, changing the structural
parameters of the rhombohedral crystal structure, is shown to be
important in determining this ordering and the resulting topological
character.
This article got published in Physical Review B: https://journals.aps.org/prb/abstract/10.1103/PhysRevB.104.035127
\end{abstract}

\maketitle

\section{Introduction}
Due to its unique electronic properties bismuth is a well studied
material and the discovery of topological phases of matter has further
motivated continued interest. Without question Bi can be considered an
important building block for topologically non-trivial compound materials
such as Bi$_2$Te$_3$ which is due to the strong spin-orbit coupling inherent
to such a heavy element. Nevertheless, the intricate details of the band
structure of bulk bismuth around the Fermi level still pose a challenge
for its accurate theoretical description. It is these details however
which are substantial for the material's properties, e.g.~the conductivity,
as well as the topology of the band structure and its respective implications.
Likewise, experiments designed to investigate the topology of bismuth are
limited by their energy resolution since a precision on the scale of only
a few \si{\milli\electronvolt} is required. As a consequence it is not
surprising that different conclusions for the topology of bismuth have
been presented in the literature. In this paper we want to contribute
to the discussion by using state of the art electronic structure methods,
including G$_0$W$_0$, which we compare to previous results. We investigate
the band inversions in the bulk via a direct comparison with the state
ordering in the atomic limit and investigate the influence of structural
parameters on the topology of the material.

Strong spin-orbit coupling such as in Bi is considered to be essential
for topologically non-trivial materials. Thus, the issue of Bi bulk topology
was addressed early on in the context of band topology \cite{fu2007, fu2007-2}
and the material was found to be either a weak topological insulator or,
based on the popular tight binding model by Liu and Allen \cite{liu1995},
topologically trivial. Although the small direct gap at the L point
(\SIrange{10}{15}{\milli\electronvolt}
\cite{aguilera2015,maltz1970,vecchi1974,isaacson1969b,brown1963,smith1964})
is by design very well reproduced by this model, the ordering and character
of the bands may not necessarily be correct. As was pointed out by Fukui and
Hatsugai \cite{fukui2007}, and later also in Refs.~\cite{ohtsubo2013,ohtsubo2016},
small changes in some of the tight binding parameters can have substantial
impact on the band topology of bulk bismuth.

Density functional theory (DFT) calculations in general agree in so far
as they predict bismuth to be a topologically trivial material, see
e.g.~\cite{hirahara2012,aguilera2015}.
Nevertheless, the underestimation of the band gap which is inherent to
DFT may be particularly problematic in the context of small direct band
gaps and the possibility of band inversions. Thus, higher levels of
theory including many-body corrections should where possible be used in
order to obtain more reliable results. To the best of our knowledge, to
date only Aguilera et al.~investigated the topology of bulk bismuth with
the generally more accurate GW quasiparticle method in Ref.~\cite{aguilera2015}.

Experimental studies are primarily restricted to the surface of bismuth
crystals or thin films and angle-resolved photoelectron spectroscopy (ARPES)
has been extensively employed. As is explained in Ref.~\cite{ito2016}
it can be concluded from the connection of the surface states on the
Bi\hkl(1 1 1) surface to the projected bulk valence or conduction bands
at $\overline{\Gamma}$ and $\overline{\text{M}}$ whether Bi is topologically trivial or non-trivial.
Since the bulk L point projects to the $\overline{\text{M}}$ point in the two-dimensional surface
Brillouin zone, an energy resolution of better than
\SI{10}{\milli\electronvolt} would also be required for reliable experimental
results if the measurement was done on a semi-infinite surface.
This precision is not easily achieved.
In thin films on the other hand, quantum confinement increases the splitting
of the states at $\overline{\text{M}}$ which at first glance further complicates the situation
\footnote{
In fact, one of the two surface states is pushed into the conduction band so
that the dispersion of the surface states on thin films consistently implies
a non-trivial topology of bulk bismuth in experiment and theory. This result
is only directly in contradiction to the calculated to\-po\-lo\-gi\-cal properties
of the bulk if the effect of confinement is not accounted for in the analysis.
}.
Artifacts due to confinement have to be removed for a thorough analysis.
As proposed in Ref.~\cite{ohtsubo2016} and measured in Refs.~\cite{ito2016,ito2020}
or used in the analysis of Ref.~\cite{chang2019}, the enhanced splitting
at $\overline{\text{M}}$ as a function of the film thickness can not only be measured with
much better overall accuracy, but it can also be exploited in order to
determine the true to\-po\-lo\-gy of the bulk. This is done via
extrapolation to the semi-infinite bulk, i.e.~practically films with
infinite thickness where the influence of confinement vanishes. These
high accuracy measurements which account for confinement have suggested
that bulk bismuth is topologically non-trivial.

We note that the topology discussed here refers to that of the three-dimensional
system and the topological classification has therefore consequences for two-dimensional
surface states. We do not discuss if bismuth is a two-dimensional topological insulator
with topological edge states and we refer to e.g.~Ref.~\cite{liu2011}.
For clarity we also point out that relatively recently higher order topology has been
found in Bi (Ref.~\cite{schindler2018}, combination of experiment and density functional
theory) as well as it has been predicted to be a first order topological crystalline insulator
(see the DFT study in Ref.~\cite{hsu2019} and Refs.~\cite{tang2019,zhang2019}).

Regardless of its band topology, the small direct gap at the L point makes Bi prone
to inversions. Therefore, different ways of inducing a topological phase transition
have been investigated. This foremost includes the effect of strain
\cite{hirahara2012,ohtsubo2013,aguilera2015,ohtsubo2016,chang2019,abdelbarey2020}
and doping \cite{jin2020}. In two dimensions it has also been shown that quantum
confinement can induce a Berry phase \cite{moore2010} and that a phase transition
can be driven via the interaction with an electric field \cite{sawahata2018}.
The effect of strain has to be considered particularly for applications since thin
films can be grown with moderate strain on mismatched substrates. In this paper,
we expand on the existing discussion of the topology of the Bi bands, using an
accurate G$_0$W$_0$ quasiparticle method to calculate the band structure and
explicitly considering the effects of strain relevant to epitaxially grown material.
We find that a transition in the ordering of states at the L-point occurs with
relatively small changes of the lattice parameters and that strain could be used
to alter the topological character of the bands.

\begin{figure*} 
  \center
  \includegraphics{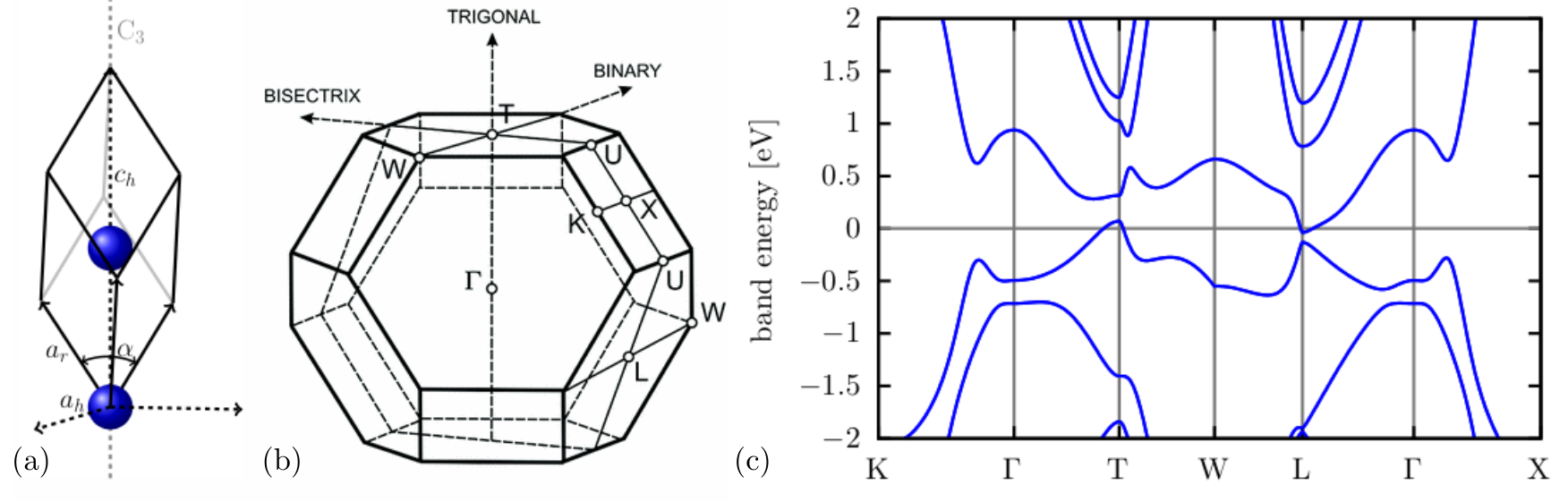}
  \caption{ \label{fig:bulkDFTBands}
  (a) The rhombohedral unit cell of Bi is defined by the cell parameters $a_r=\SI{4.7458}{\angstrom}$
  and $\alpha=\SI{57.230}{\degree}$ and the position of the second basis atom at $(1-2u) \cdot (1,1,1)$
  with $u=0.23389$ (parameters as per Ref.~\cite{schiferl1969} and used in Ref.~\cite{aguilera2015}).
  The material forms bilayers perpendicular to the trigonal \hkl[1 1 1] axis.
  (b) shows the corresponding Brillouin zone with the high symmetry points. The time reversal invariant
  momenta $\Gamma$, T, L, and X are a subset of the high symmetry points. There is only one T point and
  three inequivalent L and X points. The bulk bands for this geometry in (c) were calculated with DFT.
  In order to avoid smearing effects in the visual presentation, the Fermi level was re-calculated with the tetrahedron method
  and set to \SI{0}{\electronvolt}; the T hole pocket and L electron pockets are thus clearly visible.
  The gap and overlap are overestimated in DFT as
  discussed in the text.
  Figure (b) is reprinted with permission from Ref.~\cite{timrov2012}. Copyright (2012) by the American Physical Society.
}
\end{figure*}

\section{Computational Details}
The \textsc{Quantum Espresso} software package \cite{giannozzi2009,giannozzi2017} was used for the
density functional theory (DFT) calculations. Spin-orbit coupling (SOC) was included which requires
fully-relativistic pseudopotentials. The SG15 pseudopotentials \cite{hamann2013,schlipf2015,scherpelz2016}
were used since they are norm-conserving and thus compatible with \textsc{Yambo}. Fifteen valence electrons
(5d$^{10}$6s$^2$6p$^3$) per atom remain to be treated explicitly. We employed the generalized
gradient approximation (GGA) with the functional of Perdew, Burke, and Ernzerhof \cite{perdew1996}
as it is superior to the localized density approximation (LDA) regarding the description of the direct
gap at L and the indirect overlap between the valence band at T and the conduction band at L \cite{aguilera2015}.
A kinetic energy cutoff of \SI{50}{\rydberg} and a uniform $8 \times 8 \times 8$ $\mathbf{k}$-point grid in the Brillouin zone proved
to be sufficiently accurate for the calculation of the DFT bands.

Many-body interactions were calculated with single shot GW (G$_0$W$_0$) as it is implemented in \textsc{Yambo} \cite{marini2009,sangalli2019}.
The code treats SOC in DFT and GW in a consistent manner, i.e.~in each calculation step (see also Ref.~\cite{sakuma2011}), and we do thus not rely on the common (perturbative) post-processing treatment of SOC.
The quasiparticle equation is solved using standard approximations, i.e.~the random phase approximation (RPA) and plasmon pole approximation (PPA).
A terminator \cite{bruneval2008} reduces the number of unoccupied states required for the convergence of the correlation part of the self-energy.
No terminator was used for the screening as it turned out to introduce an unacceptable systematic error in the calculated gap and furthermore was detrimental for $\mathbf{k}$-point convergence.
Therefore, many unoccupied bands which are used as input to G$_0$W$_0$ had to be calculated in DFT which required a larger energy cutoff (\SI{100}{\rydberg}) for the DFT electronic states.

In order to plot the band structure along a path in the Brillouin zone, the G$_0$W$_0$-corrected energies which were calculated on a grid were smoothly interpolated.
This was achieved with \textsc{BoltzTraP2} \cite{madsen2006,madsen2018} which exactly reproduces the band energies on the grid points.
This includes the main high symmetry points T and L. 
\footnote{We note that the cost of exactly fitting the energies at the grid points is that minor Gibbs oscillations occur in the band energies between grid points.}

\section{DFT study: atomic limit}
A non-trivial topology of crystalline materials is associated with band inversions with respect to
the atomic limit \cite{fu2007-2,hasan2010}. The interaction between the atoms changes the re\-la\-tive
energies between the electronic states. In extreme cases this leads to a re-ordering of the states
so that a number of occupied atomic states form part of the crystal's conduction band while previously
unoccupied states drop in energy and become part of the valence band. For this to occur, the gap between
the valence and conduction bands has to close and re-open \cite{hasan2010}.
Therefore, naturally, those points in the Brillouin zone where the local gap is small are the most interesting.
For bismuth this is the L point which -- in conjunction with the overlap between the valence band at the T
point and the conduction band at L -- also determines the overall electronic properties of the material.
Fig.~\ref{fig:bulkDFTBands} shows a plot of the DFT band structure along a path between the high symmetry points.
In this section we investigate the transition from the atomic limit to the bulk material which offers
a direct way to observe bulk inversions.

Before we begin with the analysis, there are two important points to highlight regarding the theoretical framework.
First of all, bulk bismuth is a semimetal due to the indirect overlap between the valence and conduction bands.
Therefore it can not be a topological \enquote{insulator} in the conventional sense.
Nevertheless, due to the local gap everywhere in the Brillouin zone, the same formalism as for insulators applies.
As is outlined in Ref.~\cite{ohtsubo2013}, a continuous transformation corresponding to a $\mathbf{k}$-dependent
potential can be used to locally adjust the Fermi level in reciprocal space without closing the band gap.
Since the closing of a gap during the transformation is essential for a topological transition to occur, this
clearly elucidates the similarity with insulating materials and shows that a classification in terms of topologically
trivial or non-trivial phases is possible. Furthermore, Fu and Kane have shown in Ref.~\cite{fu2007-2} that the
calculation of the $\mathbb{Z}_2$ invariants is much simplified for systems like bulk bismuth which have inversion symmetry.
In this case an analysis of the valence state parities at the time reversal invariant momenta (TRIM) is sufficient in order
to calculate the four invariants $\left(\nu_0; \nu_1, \nu_2, \nu_3\right)$ that determine the topology.\footnote{Further details will be given below
but it is already apparent how band inversions play a role here since they turn valence bands into conduction bands and vice versa.}

\begin{figure*} 
\center
\includegraphics{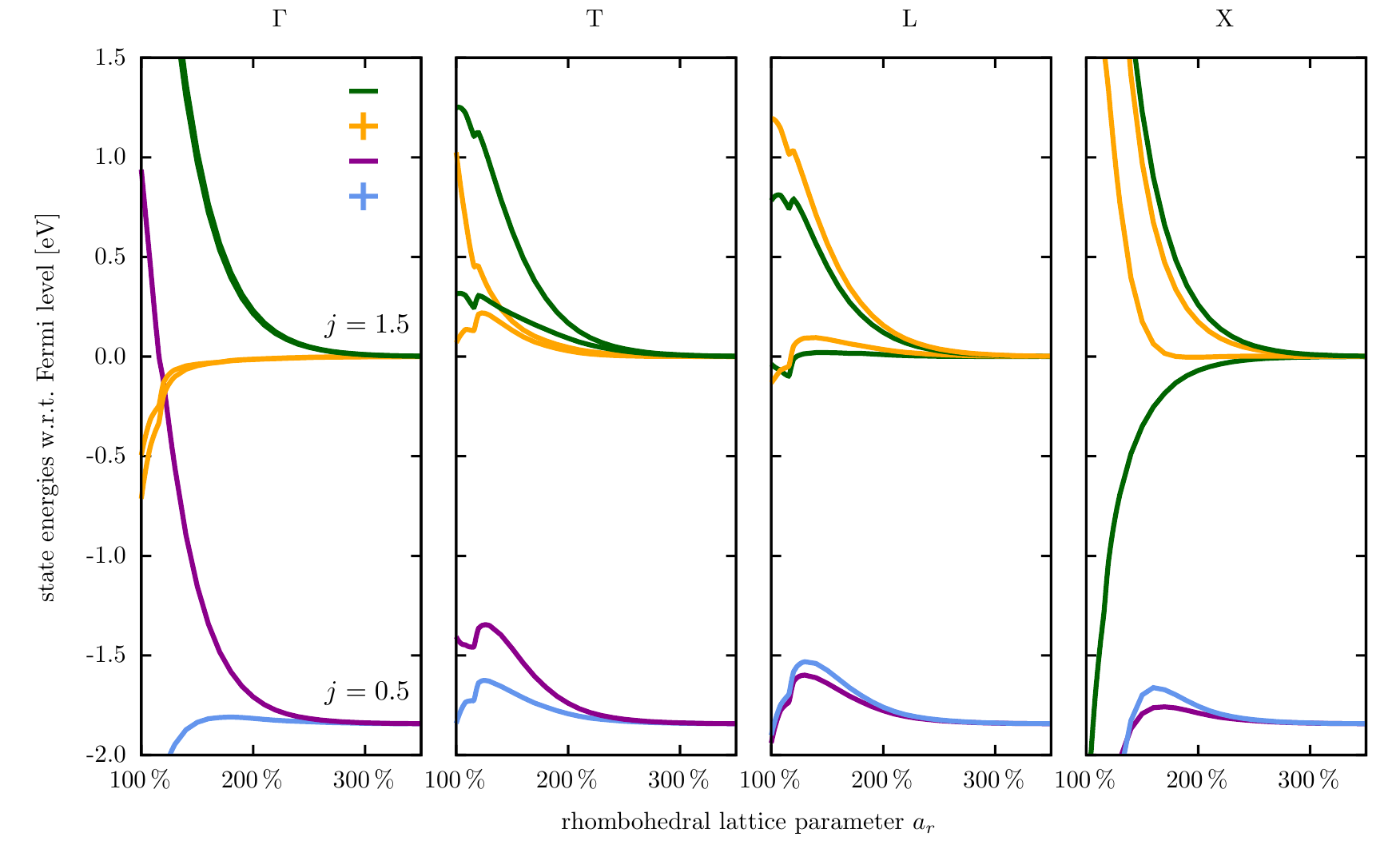}
\caption{ \label{fig:atomicLimitSoc}
  The band energies near the Fermi level in Bi versus scaling of the rhombohedral lattice parameter $a_r$. The color code implies
  the parity of the state with respect to the inversion point between the bilayers (see the legend in the
  $\Gamma$ panel). The lattice parameter $a_r$ is defined with respect to the value in
  Ref.~\cite{aguilera2015}. The internal displacement of the second atom and the angle $\alpha$ between the rhombohedral lattice vectors were not
  changed. The right hand side of each panel corresponds to the atomic limit with negligible interaction between the
  atoms. 
  The Fermi level is re-adjusted for each $a_r$ and the significant changes in the full band structure (see supplemental material \cite{supplemental-bands}) occasionally lead to kinks in the lines.
  Each line corresponds to two degenerate states (time reversal and inversion symmetry leading to Kramers
  degeneracy everywhere in the Brillouin zone).
}
\end{figure*}

Figure \ref{fig:atomicLimitSoc} shows the band energies at the four TRIM points of the rhombohedral unit cell as function of the cell parameter $a_r$ which varies
from \SI{100}{\percent} to \SI{350}{\percent}. This corresponds to an increase in volume without otherwise changing the cell geometry.
For values of the lattice parameter $a_r$ larger than three times the experimental value, the electronic state energies do not change
significantly so that Fig.~\ref{fig:atomicLimitSoc} clearly represents the transition from the bulk limit to the atomic limit. The electron configuration of
Bi is [Xe]4f$^{14}$5d$^{10}$6s$^2$6p$^3$ so that the bulk bands
at the Fermi level can be identified with the 6p states on the right hand side of each panel in Fig.~\ref{fig:atomicLimitSoc}.
We observe eight degenerate atomic states directly at the Fermi level which are the four states with $l=1$ and total angular momentum
$j=1.5$ for each of the two atoms in the unit cell. Only two of these bands are valence bands (see, e.g., the X point in the band
structure).
We distinguish valence and conduction bands by the splitting close to the atomic limit.
The remaining four p bands with $j=0.5$ lie lower in energy and are separated from the $j=1.5$ states by about
\SI{1.84}{\electronvolt} due to the strong spin-orbit coupling.

In total there are six occupied and unoccupied states in the
relevant energy range (states 25-30 and 31-36) which all originate from the 6p states of an isolated Bi atom.
We can see that inversions between the valence and conduction bands happen at the $\Gamma$ point and the three
inequivalent L points. At $\Gamma$ the purple states first cross the orange valence states when approaching
the bulk limit before the actual inversion takes place by crossing the orange conduction states. At L the
highest valence band directly inverts with the lowest conduction band. At the remaining TRIM points the order
of the states only changes within the subset of the valence bands (X) or conduction bands (T) which does not
have an impact on the topology of the material.
Clearly, the interaction at the $\Gamma$ point pushes one degenerate set of states with $j=0.5$
to higher energies so that it becomes a conduction band while in exchange two additional $j=1.5$ states now form
valence bands. Another inversion happens during the deformation of the cell at the three non-equivalent L points,
this time bet\-ween the highest valence and lowest conduction bands. The inversion happens very close to the bulk
geometry, and the energy splitting (i.e.~the L gap) remains small. This emphasizes again that an accurate theoretical
description of the system is required in order to get reliable results and we can expect the L point to have the main
impact on the topological classification of bismuth.

Overall, the even number of inversions in the bulk as calculated with DFT implies a trivial topology. This can be
checked with the method of Fu and Kane \cite{fu2007-2} and we find that for calculations of the topological invariants via
\begin{equation}
  (-1)^{\nu_0} = \prod_{\Gamma_i \in \{ \text{TRIM} \} } \delta_i ~,
\end{equation}
and the corresponding subsets of the TRIM for $\nu_1$, $\nu_2$, and $\nu_3$, considering the $2N$ occupied states
\footnote{
$\nu_1$: L$_1$, X$_1$, X$_3$, T$_1$;
$\nu_2$: L$_2$, X$_1$, X$_2$, T$_1$;
$\nu_3$: L$_3$, X$_2$, X$_3$, T$_1$;
Kramers degenerate states are considered only once.
}
in
\begin{equation}
  \delta_i     = \prod_{m=1}^{N} \xi_{2m} \left( \Gamma_i \right) ~,
  \label{eq:deltai}
\end{equation}
via their parities
\begin{equation}
  \xi_{2m} \left( \Gamma_i \right) = \pm 1 ~,
\end{equation}
only the 6p-derived bands are relevant and the product of all lower lying bands yields $+1$ for each TRIM point.
This is consistent with the argument that the core states remain in the atomic limit and also with Eq.~(\ref{eq:deltai})
which is not changed by a reordering of the valence bands only.
As can be directly deduced from the information given in Fig.~\ref{fig:atomicLimitSoc}, we find the trivial case
$\left(\nu_0; \nu_1, \nu_2, \nu_3\right) = (0;0,0,0)$ in agreement with the simple analysis of bulk inversions.
This result agrees well with all DFT calculations in the literature that we are aware of and for a more detailed
comparison we refer to the discussion below.

\begin{figure*} 
\center
\includegraphics{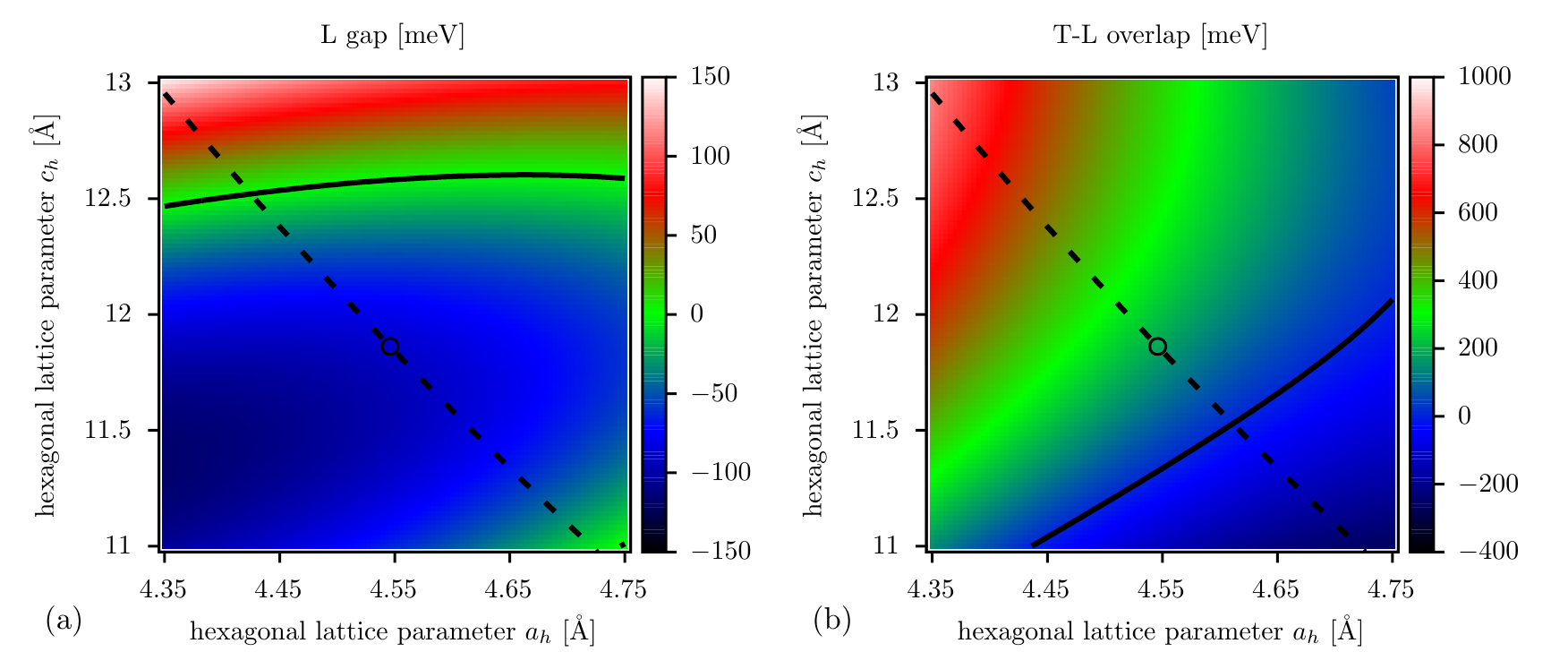}
\caption{ \label{fig:fitDFTGapAndOverlap}
  Plot (a) shows a quadratic fit of the L gap as a function of the hexagonal lattice parameters $a_h$ and $c_h$
  as calculated with DFT (GGA). The position of the second atom in the corresponding rhombohedral unit cell was
  determined by fitting the minimum energy with respect to this parameter. The experimental lattice parameters
  are marked by a circle and the line of constant volume is drawn as a dashed line. We use the convention that
  the L gap is defined to be negative for an inversion with respect to the atomic limit.
  We find that the L gap varies only slightly with $a_h$ but changes substantially with $c_h$. A topological
  phase transition occurs (only) for large $c_h$ where the L gap changes sign (solid black line) and the material
  becomes non-trivial.
  In (b) the value of the indirect overlap defined as the energy difference between the valence band at T and
  the conduction band at L is shown (quadratic fit). Here, the semimetallic characteristics of bulk bismuth
  correspond to a positive value and overlapping bands.
  The solid black line in (b) marks where the T-L band overlap vanishes.
}
\end{figure*}

\section{DFT dependence on geometry}
In the preceding section we have analyzed the band topology of bulk bismuth in its experimental equilibrium geometry.
Naturally, the changes in volume which are required to reach the atomic limit are significant.
However, the band inversion at the L point happens relatively close to the bulk limit and there are two scenarios where
this becomes important. Firstly, a strain induced topological transition of the bulk has been discussed by several
authors \cite{hirahara2012,ohtsubo2013,aguilera2015,ohtsubo2016,chang2019,abdelbarey2020,jin2020}.
In particular this applies to the growth of bismuth films on a substrate which dictates the in-plane lattice constant.
At the maximum a few percent of strain can be achieved. The second case which has to be considered are calculations
where experimental structure parameters are not used but where the crystal is relaxed so that (ideally) the lattice parameters corresponding to the theoretical energy
minimum are used. The difference between theoretical and experimental lattice parameters depends on the accuracy of the theory in general, e.g.~the
use of the LDA or GGA, as well as the quality of the pseudopotential. Thus, this case requires extra consideration.

We note that a relaxation of the system is also particularly important in lower dimensional systems, e.g.~in films
which may interact with their substrate or when a passivation of the surface is required. In this case a reorganization
of the atoms is essential and an individual control of each atomic position is not feasible, therefore these studies
rely on the capabilities of the particular implementation of the theory to find the energy minimum. From our experience,
the weak chemical bonding in bismuth can be a challenge for typical relaxation algorithms. If van der Waals corrections
are included in DFT calculations, the particular choice and parameters can also have an impact
on the structure.

Based on the discussion above, in the following we systematically investigate how the bulk properties of bismuth change
as function of the crystal structure. Although it is most intuitive to think about an expanding cell in terms of a changing
rhombohedral lattice parameter $a_r$, we here use a different approach. A hexagonal supercell with the parameters $a_h$ and
$c_h$ can be constructed which contains six atoms and more clearly reflects the layered structure of the material
\footnote{
$a_h$ defines the width of the cell in-plane with the bilayers. $c_h$ is the out-of-plane coordinate and thus controls the
distance between the bilayers. Typically, thin Bi films grow so that the bilayers lie flat on the substrate. For a conversion
between the rhombohedral and hexagonal cell see for example Ref.~\cite{hofmann2006}.
}.
Thus, although the calculations are still done for the rhombohedral cell, we provide the corresponding hexagonal
pa\-ra\-me\-ters instead for ease of interpretation.

Fig.~\ref{fig:fitDFTGapAndOverlap} shows how the direct gap at the L point and the indirect T-L overlap, as calculated
in the GGA, depend on the cell parameters. These are the crucial ob\-ser\-va\-bles for the electronic and topological
properties. The displacement of the second atom in the unit cell was determined by fitting the minimum energy with
respect to this parameter at each point in the plot. The experimental values for $a_h$ and $c_h$ are marked in the
figure and a dashed black line shows how these two pa\-ra\-me\-ters depend on each other if the experimental volume
is kept constant, as was done in Ref.~\cite{aguilera2015}.

Clearly, a variation of $a_h$ alone does not significantly change the L gap. Note that thin bismuth films preferentially
grow along the trigonal axis and that $a_h$ is the in-plane parameter of these Bi\hkl(1 1 1) films. The range in
Fig.~\ref{fig:fitDFTGapAndOverlap} is roughly $\pm\SI{4.5}{\percent}$ so that we cover all strains which might realistically
occur. On a Si\hkl(1 1 1)--$7 \times 7$ substrate the strain is \SI{-1.3}{\percent} in the plane of the bilayers \cite{nagao2004}
(compressive strain). The strain on Bi$_2$Te$_3$\hkl(1 1 1) was reported to be as large as \SI{-3.3}{\percent} \cite{hirahara2012}.

On the other hand, in contrast to $a_h$, the distance between the bilayers has a significant impact on the L band gap. To highlight this,
a solid black line in Fig.~\ref{fig:fitDFTGapAndOverlap} marks where the sign of the L gap changes. The lifting of the bulk inversion for large values
of $c_h$ leads to a topological phase transition, with only the inversion at $\Gamma$ remaining, so that the bulk becomes
topologically non-trivial. We have chosen a relatively wide range of $c_h$ values since, due to the weak interlayer
bonds, this quantity is not always well reproduced by DFT. Furthermore, depending on the flavor of DFT used, the actual
position of the transition may move slightly and we show the results for LDA in the supplemental material \cite{supplemental-main} where we also
discuss additional details. The second plot in Fig.~\ref{fig:fitDFTGapAndOverlap} shows the indirect overlap between the
T valence band and L conduction band which is responsible for the semimetallic character of the material. In line with
Ref.~\cite{aguilera2015} we observe that this quantity can vanish and a tiny band gap exists for a narrow strain window. However, the bulk in the bottom right corner is still
semimetallic since now the L valence band lies higher in energy than the conduction band at T. Thus, we do not observe
a robust semimetal-to-semiconductor transition due to strain.

It is worth highlighting that Ref.~\cite{aguilera2015} only considers constant volume and does not distinguish between
the effects of $a_h$ and $c_h$. We can observe the same topological phase transition as Aguilera et al.~for in-plane
compression (e.g.~by a substrate) and find that the transition relies on the increasing distance between the bilayers,
which is a result of relaxation due to the assumption of constant volume.
In this context the distance is to be understood as the distance between the centers of
the respective bilayers, not the minimum distance between atoms of neighboring bilayers.

\begin{figure*} 
\center
\includegraphics{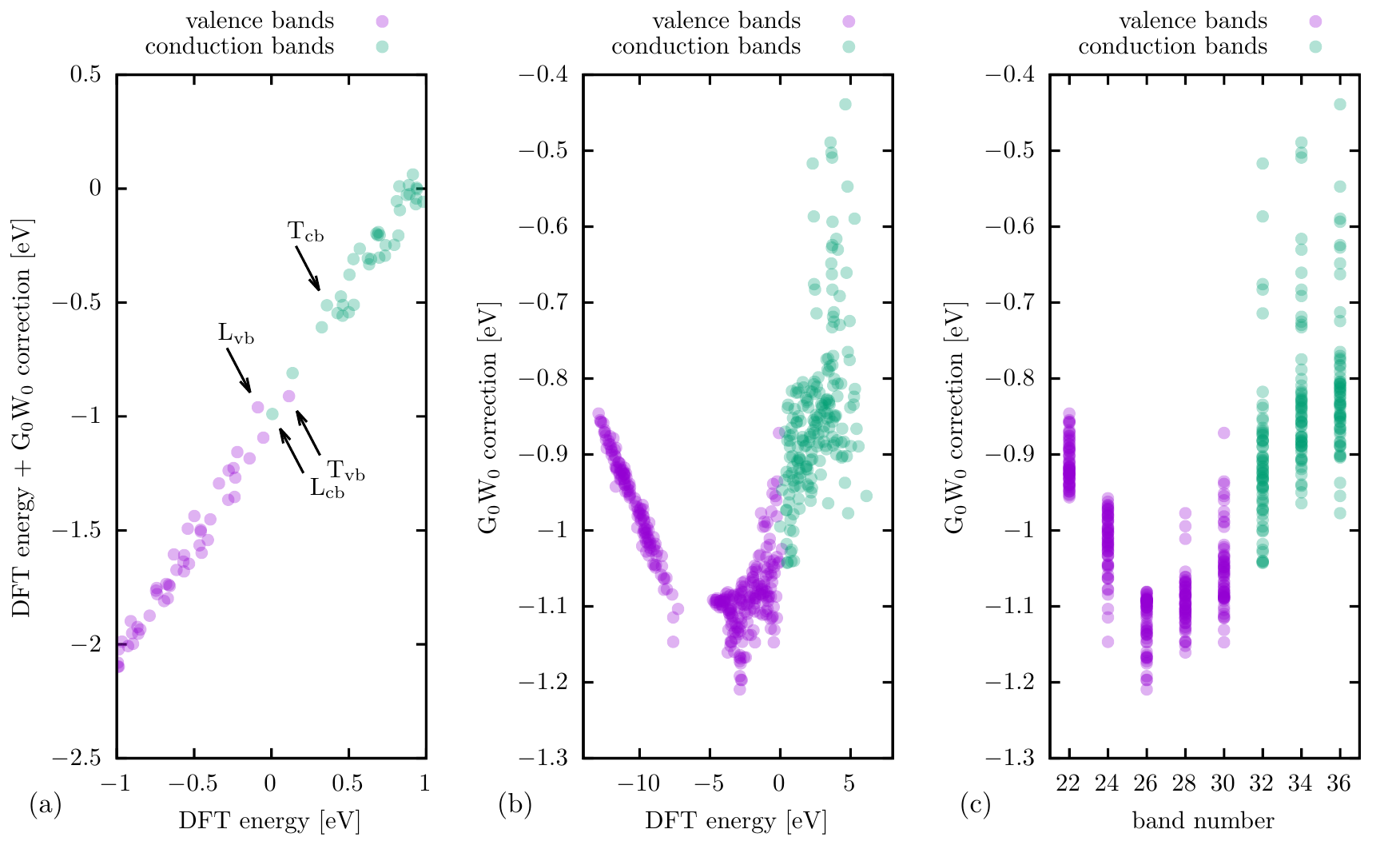}
\caption{ \label{fig:bulkCorrections}
  (a) G$_0$W$_0$ corrected state energies for the bulk cell with experimental lattice as functions of
  the DFT energy. The Fermi level of the DFT states (re-calculated by \textsc{Yambo} with smearing) is set to \SI{0}{\electronvolt}.   Labels indicate the position of the
  valence and conduction band edges at T and L.   For clarity the bare corrections are shown as function of the
  DFT energy or band in (b) and (c) to  highlight the variations within each of the bands. The effect of G$_0$W$_0$
  goes beyond a simple scissors operator which would correspond to a step-like function shifting the conduction
  bands to higher energies. The straight line for low DFT energies in (b) represents the low lying 6s states and the states above \SI{-5}{\electronvolt} in DFT energy are the 6p states.
  We note that G$_0$W$_0$ as implemented in \textsc{Yambo} does not change the wave functions so that the parities stay the same after the correction.
}
\end{figure*}

\section{Many-body corrections}
As was adequately pointed out in Ref.~\cite{aguilera2015}, the well known deficiencies of DFT demand for more
reliable methods like GW. The authors employed all-electron calculations and found that the many-body approach not only
improves the agreement between the electronic structure calculations and experiments but also confirms the DFT results
in that bismuth is topologically tri\-vi\-al. The compressive strain in-plane with the bilayers required to push the
material into the non-trivial phase was reduced to \SI{0.7}{\percent} for G$_0$W$_0$ and \SI{0.4}{\percent} for a fully
self-consistent GW, compared to DFT, which are easily achievable in experiment.

In this section we present our results for G$_0$W$_0$ as it is implemented in \textsc{Yambo}. This serves two purposes.
First of all we want to expand on our discussion of the results above and check how well DFT performs. Secondly, we
want to compare our pseudopotential based approach to that in Ref.~\cite{aguilera2015} which to the best of our knowledge
is the only available GW data for bulk bismuth. Realistically, only the computationally cheaper pseudopotential method
used here will allow for calculations of systems with many atoms such as nanostructures.
The reasons for choosing \textsc{Yambo} are outlined in the methods section.

The G$_0$W$_0$ calculations take the DFT eigenvalues as input and correct them by calculating the self-energy of the
many-body system. Therefore changes in the ordering of the states are clearly visible. Both the plasmon pole and random
phase approximation (PPA/RPA) were used in the calculations. The small size of the direct band gap demands for a high
accuracy and the calculation parameters were chosen accordingly in order to ensure adequately strict convergence for
both observables in the bulk cell, i.e.~the L gap and T-L overlap.

We find that the L gap increases from \SI{-94}{\milli\electronvolt} to \SI{29}{\milli\electronvolt}
and the indirect overlap changes from \SI{108}{\milli\electronvolt} to \SI{48}{\milli\electronvolt} (GGA compared to G$_0$W$_0$).
The significant reduction of the absolute values of both quantities compares well with Table I in Ref.~\cite{aguilera2015} and brings the calculation
in closer agreement with experiment.
The crucial difference however is that the ordering of the states at L changes and an
inversion occurs. Even without strain this inversion leads, as is evident from the discussion in the previous section,
to a topologically non-trivial bulk.
We note that the unit cell geometry used in Ref.~\cite{aguilera2015} and in our calculations is identical.

Fig.~\ref{fig:bulkCorrections} shows the effect of the many-body corrections on the DFT bands. The eigenvalues of all 6s and 6p
bands were calculated on the full $8 \times 8 \times 8$ grid. In Fig.~\ref{fig:bulkCorrections} (a) only the states in an energy
window of $\pm \SI{1}{\electronvolt}$ around the Fermi level are considered, corresponding to the first valence and conduction
band. The state energies at L and T are labelled to highlight the re-ordering at the L point. It is often assumed that the effect
of G$_0$W$_0$ can well be approximated by a rigid shift of the conduction bands with respect to the valence bands (scissors operator).
Based on the changes in the direct gap and the overlap, a shift of the order of \SI{100}{\milli\electronvolt} might be expected.
It is however clear from Fig.~\ref{fig:bulkCorrections} (a) that there is considerable variation in the corrections within
each of the bands and any small offset between the valence and conduction bands, if present, is disguised.
For clarity Fig.~\ref{fig:bulkCorrections} (b) shows the bare G$_0$W$_0$ corrections as a function of the DFT energy; Fig.~\ref{fig:bulkCorrections} (c)
shows the band resolved corrections. We conclude that a scissors operator does not represent the many-body corrections well,
which is consistent with the inversion of the valence and conduction bands at the L point
and similar findings for complex systems with strong SOC, e.g.~in Ref.~\cite{sakuma2011}.

\begin{figure} 
\center
\includegraphics{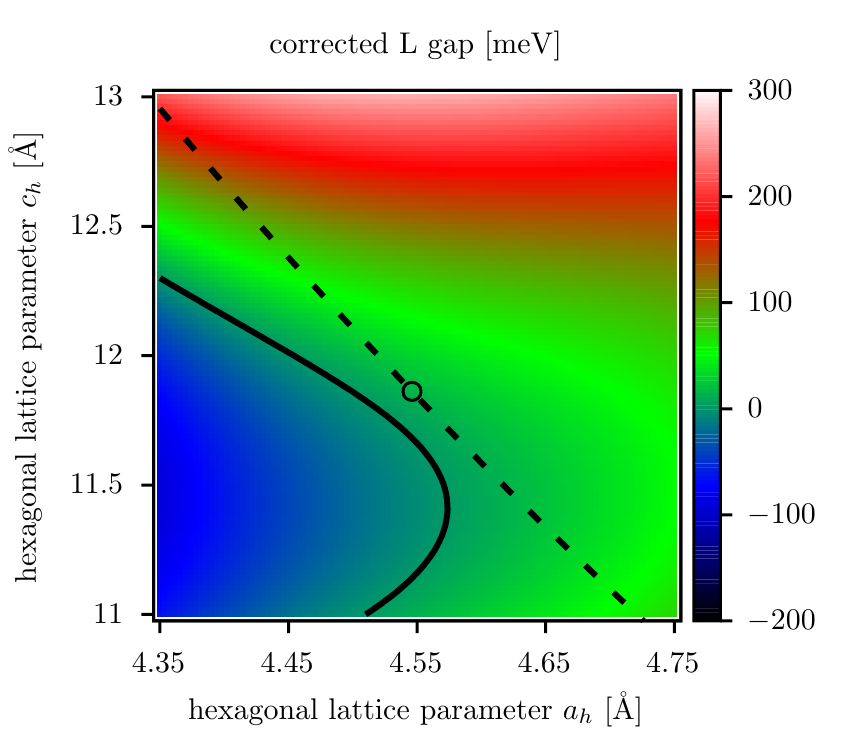}
\caption{ \label{fig:fitGWGap}
  G$_0$W$_0$ corrected L gap of Bi as function of geometry. In contrast to the DFT results in Fig.~\ref{fig:fitDFTGapAndOverlap},
  the topologically trivial region is limited to small $a_h$ and $c_h$.   In particular, cells with constant experimental volume
  are all topologically non-trivial. We note that compared to the DFT results, the effect of the parameter $a_h$ increased substantially.
  For better accuracy the corrections were fitted with a third order polynomial.
}
\end{figure}

For comparison, Fig.~\ref{fig:fitGWGap} shows the L gap including the G$_0$W$_0$ energy correction as a function of geometry. Most notably,
the topologically trivial region is no longer crossed by the line of constant volume. Therefore, within the limitations
of the theoretical framework presented above, bulk bismuth is topologically non-trivial and a compressive strain
in all directions would be required to make the material trivial.

\begin{figure*} 
\center
\includegraphics{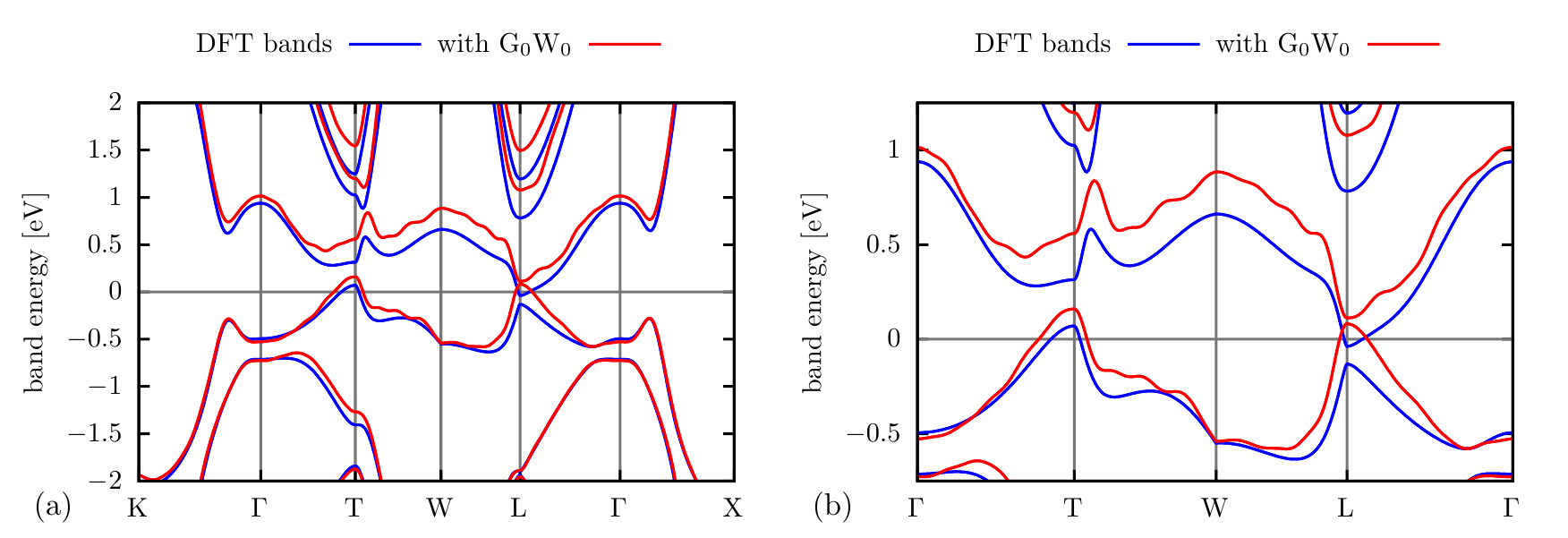}
\caption{ \label{fig:correctedBandStructure}
  Comparison of the DFT band structure as calculated in \textsc{Quantum Espresso} with the G$_0$W$_0$ results
  interpolated along the same path.
  (a) Shows bands within \SI{2}{\electronvolt} of the DFT Fermi level which was calculated with the tetrahedron method and (b) shows bands from
  \SIrange{-0.75}{+1.25}{\electronvolt} of the DFT Fermi level to highlight the band structure near L, T, and $\Gamma$.
  A $16 \times 16 \times 16$ grid is sufficient to well reproduce the \textsc{Quantum Espresso} bands via an interpolation
  of the DFT states. G$_0$W$_0$ calculations on such a dense grid are however very expensive. Thus, we interpolated the
  quasiparticle corrections from the $8 \times 8 \times 8$ grid to a $16 \times 16 \times 16$ grid which was then combined
  with the DFT energies before the final interpolation along the path. The energies were shifted so that the valence bands
  approximately align with the DFT bands.
  Using \textsc{BoltzTraP2} ensures that the state energies at $\Gamma$, T, L, and X are exactly reproduced
  in the plot although the method introduces, depending on the parameters, some or many Gibbs oscillations between the
  high symmetry points (unphysical artifact of the method).
}
\end{figure*}

Finally, in Fig.~\ref{fig:correctedBandStructure} we show a comparison of the band structures before and after correction.
The T, W, and L points are most affected by the many-body approach.
The re-ordering of the states was considered prior to the interpolation of the bands with \textsc{BoltzTraP2}, to reproduce the expected avoided crossing \cite{aguadopuente2020} without explicitly re-diagonalizing the Hamiltonian.

\section{Discussion}
In the first part of this paper we discussed the nature of the band inversions in bulk bismuth. By comparing with
the atomic limit, we have shown that within DFT there is an inversion at $\Gamma$ and at each of the three L points.
The even number of inversions indicates a topologically trivial material which is consistent with our analysis of
the parities at the TRIM points and the DFT results in the literature. We note that our convention for defining the parities at T and L differs
from the calculations in Refs.~\cite{liu1995,golin1968} (tight binding and pseudopotential, respectively) as well
as from the \textsc{Vasp} DFT calculations in Refs.~\cite{hirahara2012,schindler2018,hsu2019,jin2020}.
This is simply due to the fact that we have defined the parity with respect to the inversion point between the bilayers.
If we instead choose the inversion point within the bilayers, then the parities at T and L change sign and we recover those
of the references above. The analysis and results remain the same, as expected since the topology should not depend on the
choice of the inversion center. As was pointed out in Ref.~\cite{ohtsubo2013}, there are no measurements of the state
parities which could be used as a benchmark and experimental papers often use the symmetry labels which were obtained by one
of the theoretical works above. By tracking the evolution of the states from the atomic limit to the bulk lattice, a much more
concise picture of the bulk inversions was obtained compared to the usual comparison of the parities alone. In particular
this method allows us to reliably distinguish inversions between valence and conduction bands from those within each
individual group of bands. The importance of this distinction becomes clear at $\Gamma$.
In agreement with Ref.~\cite{hsu2019}, we there find three states with even parity below the Fermi level which is however
related to just a single inversion of valence and conduction bands.

Our G$_0$W$_0$ calculations indicate that Bi is a topologically non-trivial material.
This is in line with recent high accuracy measurements in Refs.~\cite{ohtsubo2013,ito2016,ito2020}
but does not agree with the results of Aguilera et al.~\cite{aguilera2015}
where a small compressive strain in the bilayer plane was required to induce a transition
into the non-trivial regime.
We are not aware of any other GW calculations for bismuth in the literature which we could compare our results to.
While SOC, which is critical for the inversion, was treated in each step of both calculations, we
identify four main differences. The authors of Ref.~\cite{aguilera2015} employed all-electron
calculations while our calculations are based on pseudopotentials. Also, only our calculations use the PPA.
Another difference is that Ref.~\cite{aguilera2015} uses the full self-energy matrix (not just
the diagonal elements) and updates the wave functions in the single-shot calculation.
Furthermore, the internal displacement of the basis atoms in the unit cell was kept constant
in Ref.~\cite{aguilera2015} while we fitted the optimum atomic positions when applying strain.
We note that the absolute value of the gap and overlap which are the common observables agree
well in the absence of strain. Overall, we conclude that even with these high accuracy methods
no unambiguous claim can be made with regards to the topological properties of bismuth.
The calculated L gap appears to be very sensitive to the calculation details.

The strain induced topological phase transition discussed by several authors (see above)
usually considers the effect of in-plane strain.
The importance of the interbilayer distance has been pointed out in Ref.~\cite{liu2011}
for the two-dimensional system.
The effect on the bulk has been shown in an extreme case in Ref.~\cite{chang2019}
and has also been hinted at in Ref.~\cite{jin2020}.
Overall, different publications made different assumptions regarding the crystal structure,
e.g.~a constant volume and fixed $u$ in Ref.~\cite{aguilera2015},
fixed $a_h$ and changing $c_h$ while keeping $u$ fixed in Ref.~\cite{chang2019}
or an overall expansion of the cell in Ref.~\cite{jin2020}.
Avoiding any issues related to the relaxation with weak interlayer bonds, we have expanded this discussion and
sys\-te\-ma\-ti\-cal\-ly investigated the effect of the lattice geometry on the L gap (and overlap).
Since there is no direct experimental control over the internal displacement parameter,
it was determined by fitting the position of lowest energy.
We believe that our discussion above provides a more complete picture of the topological
phases of bulk bismuth and is particularly useful for the interpretation of experimental data.
From Fig.~\ref{fig:fitDFTGapAndOverlap} we conclude that the often considered
in-plane strain only indirectly gives rise to a topological phase transition by effectively
increasing the distance between the bilayers (in line with the assumption of constant volume
in Ref.~\cite{aguilera2015}).
Our DFT calculations show that in comparison to the in-plane lattice constant, the bilayer distance
is a very important parameter which may not have been expected, given the weak interlayer bonds.
Notably, under the not unreasonable approximation of constant volume,
the trivial phase can no longer be reached in our G$_0$W$_0$ calculation, see Fig.~\ref{fig:fitGWGap},
i.e.~the non-trivial phase is stable with respect to strain.

Due to the small L gap, differences in the band to\-po\-lo\-gy between the various calculations are
always likely to be the result of a discrepancy at the L points.
The G$_0$W$_0$ many-body corrections remove the inversion at L in the bulk, similar to the
modified tight binding models mentioned in the introduction.
In general, the valence and conduction states at the other TRIM points are sufficiently separated
in energy so that they should not give rise to controversy.
In the context of two-dimensional systems,
the apparent non-trivial surface state dispersion of thin Bi\hkl(1 1 1) films, which is the result
of an increased splitting of the valence and conduction bands due to confinement, is well
reproduced even with DFT. Thus, the authors of Ref.~\cite{ohtsubo2013} assume that confinement
lifts the -- presumably artificial -- DFT L point bulk inversions.
We note that in this case it is not as straightforward to construct a corresponding three-dimensional system
as it is for strained films.

Finally, we want to highlight that knowledge about the dependence of the topology,
L gap and overlap on the crystal structure is very valuable in the context of DFT calculations.
Often the experimental lattice pa\-ra\-me\-ters can not be used.
This is for example the case for nanostructures and at interfaces where the atoms
have to be relaxed and control over each atomic coordinate is not feasible.
In particular, the weak bonds between the bilayers are a challenge, i.e.~the distance
between the bilayers can deviate substantially from the experimental value even in the bulk
so that $c_h$ and the unit cell volume are affected.
Accordingly, the side effects on the crucial details of the band structure
should be considered in the interpretation of the results.

\section{Conclusions}
All in all, our results are in agreement with recent experiments which predict a non-trivial bulk material.
The band gap at L is in much better agreement with respect to experiment after calculation of the G$_0$W$_0$ correction.
For constant volume deformations, the band topology does not change.
The topologically trivial region is limited to small values for the lattice parameters $a_h$ and $c_h$ within the G$_0$W$_0$ calculations.
For measurements on \hkl(1 1 1) free standing films, it is anticipated that compressive strain for $a_h$ will result in strain relaxation in the $c_h$ lattice parameter, hence this region may not be easily accessible to experiment.
Compared to the in-plane lattice parameter $a_h$, the out-of-plane lattice parameter $c_h$ has an unexpectedly high impact on the L gap -- this effect is larger in the DFT results than as predicted from G$_0$W$_0$ but it is a notable effect seen in both methods.

\begin{acknowledgments}
\section{Acknowledgments}
This work has been funded by Science Foundation Ireland through the Principal Investigator Award No.~13/IA/1956.
The authors wish to acknowledge the Irish Centre for High-End Computing (ICHEC) for the provision of computational facilities and support.
Support is also provided by the Nottingham Ningbo New Materials Institute and the National Natural Science Foundation of China with Project Code 61974079.
\end{acknowledgments}

\bibliography{./bibliography}

\begin{thebibliography}{51}%
\makeatletter
\providecommand \@ifxundefined [1]{%
 \@ifx{#1\undefined}
}%
\providecommand \@ifnum [1]{%
 \ifnum #1\expandafter \@firstoftwo
 \else \expandafter \@secondoftwo
 \fi
}%
\providecommand \@ifx [1]{%
 \ifx #1\expandafter \@firstoftwo
 \else \expandafter \@secondoftwo
 \fi
}%
\providecommand \natexlab [1]{#1}%
\providecommand \enquote  [1]{``#1''}%
\providecommand \bibnamefont  [1]{#1}%
\providecommand \bibfnamefont [1]{#1}%
\providecommand \citenamefont [1]{#1}%
\providecommand \href@noop [0]{\@secondoftwo}%
\providecommand \href [0]{\begingroup \@sanitize@url \@href}%
\providecommand \@href[1]{\@@startlink{#1}\@@href}%
\providecommand \@@href[1]{\endgroup#1\@@endlink}%
\providecommand \@sanitize@url [0]{\catcode `\\12\catcode `\$12\catcode
  `\&12\catcode `\#12\catcode `\^12\catcode `\_12\catcode `\%12\relax}%
\providecommand \@@startlink[1]{}%
\providecommand \@@endlink[0]{}%
\providecommand \url  [0]{\begingroup\@sanitize@url \@url }%
\providecommand \@url [1]{\endgroup\@href {#1}{\urlprefix }}%
\providecommand \urlprefix  [0]{URL }%
\providecommand \Eprint [0]{\href }%
\providecommand \doibase [0]{http://dx.doi.org/}%
\providecommand \selectlanguage [0]{\@gobble}%
\providecommand \bibinfo  [0]{\@secondoftwo}%
\providecommand \bibfield  [0]{\@secondoftwo}%
\providecommand \translation [1]{[#1]}%
\providecommand \BibitemOpen [0]{}%
\providecommand \bibitemStop [0]{}%
\providecommand \bibitemNoStop [0]{.\EOS\space}%
\providecommand \EOS [0]{\spacefactor3000\relax}%
\providecommand \BibitemShut  [1]{\csname bibitem#1\endcsname}%
\let\auto@bib@innerbib\@empty
\bibitem [{\citenamefont {Fu}\ \emph {et~al.}(2007)\citenamefont {Fu},
  \citenamefont {Kane},\ and\ \citenamefont {Mele}}]{fu2007}%
  \BibitemOpen
  \bibfield  {author} {\bibinfo {author} {\bibfnamefont {L.}~\bibnamefont
  {Fu}}, \bibinfo {author} {\bibfnamefont {C.~L.}\ \bibnamefont {Kane}}, \ and\
  \bibinfo {author} {\bibfnamefont {E.~J.}\ \bibnamefont {Mele}},\ }\bibfield
  {title} {\enquote {\bibinfo {title} {Topological insulators in three
  dimensions},}\ }\href@noop {} {\bibfield  {journal} {\bibinfo  {journal}
  {Phys. Rev. Lett.}\ }\textbf {\bibinfo {volume} {98}},\ \bibinfo {pages}
  {106803} (\bibinfo {year} {2007})}\BibitemShut {NoStop}%
\bibitem [{\citenamefont {Fu}\ and\ \citenamefont {Kane}(2007)}]{fu2007-2}%
  \BibitemOpen
  \bibfield  {author} {\bibinfo {author} {\bibfnamefont {L.}~\bibnamefont
  {Fu}}\ and\ \bibinfo {author} {\bibfnamefont {C.~L.}\ \bibnamefont {Kane}},\
  }\bibfield  {title} {\enquote {\bibinfo {title} {Topological insulators with
  inversion symmetry},}\ }\href@noop {} {\bibfield  {journal} {\bibinfo
  {journal} {Phys. Rev. B}\ }\textbf {\bibinfo {volume} {76}},\ \bibinfo
  {pages} {045302} (\bibinfo {year} {2007})}\BibitemShut {NoStop}%
\bibitem [{\citenamefont {Liu}\ and\ \citenamefont {Allen}(1995)}]{liu1995}%
  \BibitemOpen
  \bibfield  {author} {\bibinfo {author} {\bibfnamefont {Y.}~\bibnamefont
  {Liu}}\ and\ \bibinfo {author} {\bibfnamefont {R.~E.}\ \bibnamefont
  {Allen}},\ }\bibfield  {title} {\enquote {\bibinfo {title} {Electronic
  structure of the semimetals {Bi} and {Sb}},}\ }\href@noop {} {\bibfield
  {journal} {\bibinfo  {journal} {Phys. Rev. B}\ }\textbf {\bibinfo {volume}
  {52}},\ \bibinfo {pages} {1566} (\bibinfo {year} {1995})}\BibitemShut
  {NoStop}%
\bibitem [{\citenamefont {Aguilera}\ \emph {et~al.}(2015)\citenamefont
  {Aguilera}, \citenamefont {Friedrich},\ and\ \citenamefont
  {Blügel}}]{aguilera2015}%
  \BibitemOpen
  \bibfield  {author} {\bibinfo {author} {\bibfnamefont {I.}~\bibnamefont
  {Aguilera}}, \bibinfo {author} {\bibfnamefont {C.}~\bibnamefont {Friedrich}},
  \ and\ \bibinfo {author} {\bibfnamefont {S.}~\bibnamefont {Blügel}},\
  }\bibfield  {title} {\enquote {\bibinfo {title} {Electronic phase transitions
  of bismuth under strain from relativistic self-consistent {GW}
  calculations},}\ }\href@noop {} {\bibfield  {journal} {\bibinfo  {journal}
  {Phys. Rev. B}\ }\textbf {\bibinfo {volume} {91}},\ \bibinfo {pages} {125129}
  (\bibinfo {year} {2015})}\BibitemShut {NoStop}%
\bibitem [{\citenamefont {Maltz}\ and\ \citenamefont
  {Dresselhaus}(1970)}]{maltz1970}%
  \BibitemOpen
  \bibfield  {author} {\bibinfo {author} {\bibfnamefont {M.}~\bibnamefont
  {Maltz}}\ and\ \bibinfo {author} {\bibfnamefont {M.~S.}\ \bibnamefont
  {Dresselhaus}},\ }\bibfield  {title} {\enquote {\bibinfo {title}
  {Magnetoreflection studies in bismuth},}\ }\href@noop {} {\bibfield
  {journal} {\bibinfo  {journal} {Phys. Rev. B}\ }\textbf {\bibinfo {volume}
  {2}},\ \bibinfo {pages} {2877} (\bibinfo {year} {1970})}\BibitemShut
  {NoStop}%
\bibitem [{\citenamefont {Vecchi}\ and\ \citenamefont
  {Dresselhaus}(1974)}]{vecchi1974}%
  \BibitemOpen
  \bibfield  {author} {\bibinfo {author} {\bibfnamefont {M.~P.}\ \bibnamefont
  {Vecchi}}\ and\ \bibinfo {author} {\bibfnamefont {M.~S.}\ \bibnamefont
  {Dresselhaus}},\ }\bibfield  {title} {\enquote {\bibinfo {title} {Temperature
  dependence of the band parameters of bismuth},}\ }\href@noop {} {\bibfield
  {journal} {\bibinfo  {journal} {Phys. Rev. B}\ }\textbf {\bibinfo {volume}
  {10}},\ \bibinfo {pages} {771} (\bibinfo {year} {1974})}\BibitemShut
  {NoStop}%
\bibitem [{\citenamefont {Isaacson}\ and\ \citenamefont
  {Williams}(1969)}]{isaacson1969b}%
  \BibitemOpen
  \bibfield  {author} {\bibinfo {author} {\bibfnamefont {R.~T.}\ \bibnamefont
  {Isaacson}}\ and\ \bibinfo {author} {\bibfnamefont {G.~A.}\ \bibnamefont
  {Williams}},\ }\bibfield  {title} {\enquote {\bibinfo {title} {Alfvén-wave
  propagation in solid-state plasmas. {III}. {Quantum} oscillations of the
  {Fermi} surface of bismuth},}\ }\href@noop {} {\bibfield  {journal} {\bibinfo
   {journal} {Phys. Rev.}\ }\textbf {\bibinfo {volume} {185}},\ \bibinfo
  {pages} {682} (\bibinfo {year} {1969})}\BibitemShut {NoStop}%
\bibitem [{\citenamefont {Brown}\ \emph {et~al.}(1963)\citenamefont {Brown},
  \citenamefont {Mavroides},\ and\ \citenamefont {Lax}}]{brown1963}%
  \BibitemOpen
  \bibfield  {author} {\bibinfo {author} {\bibfnamefont {R.~N.}\ \bibnamefont
  {Brown}}, \bibinfo {author} {\bibfnamefont {J.~G.}\ \bibnamefont
  {Mavroides}}, \ and\ \bibinfo {author} {\bibfnamefont {B.}~\bibnamefont
  {Lax}},\ }\bibfield  {title} {\enquote {\bibinfo {title} {Magnetoreflection
  in bismuth},}\ }\href@noop {} {\bibfield  {journal} {\bibinfo  {journal}
  {Phys. Rev.}\ }\textbf {\bibinfo {volume} {129}},\ \bibinfo {pages} {2055}
  (\bibinfo {year} {1963})}\BibitemShut {NoStop}%
\bibitem [{\citenamefont {Smith}\ \emph {et~al.}(1964)\citenamefont {Smith},
  \citenamefont {Baraff},\ and\ \citenamefont {Rowell}}]{smith1964}%
  \BibitemOpen
  \bibfield  {author} {\bibinfo {author} {\bibfnamefont {G.~E.}\ \bibnamefont
  {Smith}}, \bibinfo {author} {\bibfnamefont {G.~A.}\ \bibnamefont {Baraff}}, \
  and\ \bibinfo {author} {\bibfnamefont {J.~M.}\ \bibnamefont {Rowell}},\
  }\bibfield  {title} {\enquote {\bibinfo {title} {Effective $g$ factor of
  electrons and holes in bismuth},}\ }\href@noop {} {\bibfield  {journal}
  {\bibinfo  {journal} {Phys. Rev.}\ }\textbf {\bibinfo {volume} {135}},\
  \bibinfo {pages} {A1118} (\bibinfo {year} {1964})}\BibitemShut {NoStop}%
\bibitem [{\citenamefont {Fukui}\ and\ \citenamefont
  {Hatsugai}(2007)}]{fukui2007}%
  \BibitemOpen
  \bibfield  {author} {\bibinfo {author} {\bibfnamefont {T.}~\bibnamefont
  {Fukui}}\ and\ \bibinfo {author} {\bibfnamefont {Y.}~\bibnamefont
  {Hatsugai}},\ }\bibfield  {title} {\enquote {\bibinfo {title} {Quantum {S}pin
  {H}all effect in three dimensional materials: Lattice computation of {Z}$_2$
  topological invariants and its application to {Bi} and {Sb}},}\ }\href@noop
  {} {\bibfield  {journal} {\bibinfo  {journal} {J. Phys. Soc. Jpn.}\ }\textbf
  {\bibinfo {volume} {76}},\ \bibinfo {pages} {053702} (\bibinfo {year}
  {2007})}\BibitemShut {NoStop}%
\bibitem [{\citenamefont {Ohtsubo}\ \emph {et~al.}(2013)\citenamefont
  {Ohtsubo}, \citenamefont {Perfetti}, \citenamefont {Goerbig}, \citenamefont
  {Le~F{\`{e}}vre}, \citenamefont {Bertran},\ and\ \citenamefont
  {Taleb-Ibrahimi}}]{ohtsubo2013}%
  \BibitemOpen
  \bibfield  {author} {\bibinfo {author} {\bibfnamefont {Y.}~\bibnamefont
  {Ohtsubo}}, \bibinfo {author} {\bibfnamefont {L.}~\bibnamefont {Perfetti}},
  \bibinfo {author} {\bibfnamefont {M.~O.}\ \bibnamefont {Goerbig}}, \bibinfo
  {author} {\bibfnamefont {P.}~\bibnamefont {Le~F{\`{e}}vre}}, \bibinfo
  {author} {\bibfnamefont {F.}~\bibnamefont {Bertran}}, \ and\ \bibinfo
  {author} {\bibfnamefont {A.}~\bibnamefont {Taleb-Ibrahimi}},\ }\bibfield
  {title} {\enquote {\bibinfo {title} {Non-trivial surface-band dispersion on
  {Bi}\hkl(111)},}\ }\href@noop {} {\bibfield  {journal} {\bibinfo  {journal}
  {New J. Phys.}\ }\textbf {\bibinfo {volume} {15}},\ \bibinfo {pages} {033041}
  (\bibinfo {year} {2013})}\BibitemShut {NoStop}%
\bibitem [{\citenamefont {Ohtsubo}\ and\ \citenamefont
  {Kimura}(2016)}]{ohtsubo2016}%
  \BibitemOpen
  \bibfield  {author} {\bibinfo {author} {\bibfnamefont {Y.}~\bibnamefont
  {Ohtsubo}}\ and\ \bibinfo {author} {\bibfnamefont {S.}~\bibnamefont
  {Kimura}},\ }\bibfield  {title} {\enquote {\bibinfo {title} {Topological
  phase transition of single-crystal {Bi} based on empirical tight-binding
  calculations},}\ }\href@noop {} {\bibfield  {journal} {\bibinfo  {journal}
  {New J. Phys.}\ }\textbf {\bibinfo {volume} {18}},\ \bibinfo {pages} {123015}
  (\bibinfo {year} {2016})}\BibitemShut {NoStop}%
\bibitem [{\citenamefont {Hirahara}\ \emph {et~al.}(2012)\citenamefont
  {Hirahara}, \citenamefont {Fukui}, \citenamefont {Shirasawa}, \citenamefont
  {Yamada}, \citenamefont {Aitani}, \citenamefont {Miyazaki}, \citenamefont
  {Matsunami}, \citenamefont {Kimura}, \citenamefont {Takahashi}, \citenamefont
  {Hasegawa},\ and\ \citenamefont {Kobayashi}}]{hirahara2012}%
  \BibitemOpen
  \bibfield  {author} {\bibinfo {author} {\bibfnamefont {T.}~\bibnamefont
  {Hirahara}}, \bibinfo {author} {\bibfnamefont {N.}~\bibnamefont {Fukui}},
  \bibinfo {author} {\bibfnamefont {T.}~\bibnamefont {Shirasawa}}, \bibinfo
  {author} {\bibfnamefont {M.}~\bibnamefont {Yamada}}, \bibinfo {author}
  {\bibfnamefont {M.}~\bibnamefont {Aitani}}, \bibinfo {author} {\bibfnamefont
  {H.}~\bibnamefont {Miyazaki}}, \bibinfo {author} {\bibfnamefont
  {M.}~\bibnamefont {Matsunami}}, \bibinfo {author} {\bibfnamefont
  {S.}~\bibnamefont {Kimura}}, \bibinfo {author} {\bibfnamefont
  {T.}~\bibnamefont {Takahashi}}, \bibinfo {author} {\bibfnamefont
  {S.}~\bibnamefont {Hasegawa}}, \ and\ \bibinfo {author} {\bibfnamefont
  {K.}~\bibnamefont {Kobayashi}},\ }\bibfield  {title} {\enquote {\bibinfo
  {title} {Atomic and electronic structure of ultrathin {Bi}\hkl(111) films
  grown on {Bi}$_2${Te}$_3$\hkl(111) substrates: Evidence for a strain-induced
  topological phase transition},}\ }\href@noop {} {\bibfield  {journal}
  {\bibinfo  {journal} {Phys. Rev. Lett.}\ }\textbf {\bibinfo {volume} {109}},\
  \bibinfo {pages} {227401} (\bibinfo {year} {2012})}\BibitemShut {NoStop}%
\bibitem [{\citenamefont {Ito}\ \emph {et~al.}(2016)\citenamefont {Ito},
  \citenamefont {Feng}, \citenamefont {Arita}, \citenamefont {Takayama},
  \citenamefont {Liu}, \citenamefont {Someya}, \citenamefont {Chen},
  \citenamefont {Iimori}, \citenamefont {Namatame}, \citenamefont {Taniguchi},
  \citenamefont {Cheng}, \citenamefont {Tang}, \citenamefont {Komori},
  \citenamefont {Kobayashi}, \citenamefont {Chiang},\ and\ \citenamefont
  {Matsuda}}]{ito2016}%
  \BibitemOpen
  \bibfield  {author} {\bibinfo {author} {\bibfnamefont {S.}~\bibnamefont
  {Ito}}, \bibinfo {author} {\bibfnamefont {B.}~\bibnamefont {Feng}}, \bibinfo
  {author} {\bibfnamefont {M.}~\bibnamefont {Arita}}, \bibinfo {author}
  {\bibfnamefont {A.}~\bibnamefont {Takayama}}, \bibinfo {author}
  {\bibfnamefont {R.-Y.}\ \bibnamefont {Liu}}, \bibinfo {author} {\bibfnamefont
  {T.}~\bibnamefont {Someya}}, \bibinfo {author} {\bibfnamefont {W.-C.}\
  \bibnamefont {Chen}}, \bibinfo {author} {\bibfnamefont {T.}~\bibnamefont
  {Iimori}}, \bibinfo {author} {\bibfnamefont {H.}~\bibnamefont {Namatame}},
  \bibinfo {author} {\bibfnamefont {M.}~\bibnamefont {Taniguchi}}, \bibinfo
  {author} {\bibfnamefont {C.-M.}\ \bibnamefont {Cheng}}, \bibinfo {author}
  {\bibfnamefont {S.-J.}\ \bibnamefont {Tang}}, \bibinfo {author}
  {\bibfnamefont {F.}~\bibnamefont {Komori}}, \bibinfo {author} {\bibfnamefont
  {K.}~\bibnamefont {Kobayashi}}, \bibinfo {author} {\bibfnamefont {T.-C.}\
  \bibnamefont {Chiang}}, \ and\ \bibinfo {author} {\bibfnamefont
  {I.}~\bibnamefont {Matsuda}},\ }\bibfield  {title} {\enquote {\bibinfo
  {title} {Proving nontrivial topology of pure bismuth by quantum
  confinement},}\ }\href@noop {} {\bibfield  {journal} {\bibinfo  {journal}
  {Phys. Rev. Lett.}\ }\textbf {\bibinfo {volume} {117}},\ \bibinfo {pages}
  {236402} (\bibinfo {year} {2016})}\BibitemShut {NoStop}%
\bibitem [{Note1()}]{Note1}%
  \BibitemOpen
  \bibinfo {note} {In fact, one of the two surface states is pushed into the
  conduction band so that the dispersion of the surface states on thin films
  consistently implies a non-trivial topology of bulk bismuth in experiment and
  theory. This result is only directly in contradiction to the calculated
  to\protect \-po\protect \-lo\protect \-gi\protect \-cal properties of the
  bulk if the effect of confinement is not accounted for in the
  analysis.}\BibitemShut {Stop}%
\bibitem [{\citenamefont {Ito}\ \emph {et~al.}(2020)\citenamefont {Ito},
  \citenamefont {Arita}, \citenamefont {Haruyama}, \citenamefont {Feng},
  \citenamefont {Chen}, \citenamefont {Namatame}, \citenamefont {Taniguchi},
  \citenamefont {Cheng}, \citenamefont {Bian}, \citenamefont {Tang},
  \citenamefont {Chiang}, \citenamefont {Sugino}, \citenamefont {Komori},\ and\
  \citenamefont {Matsuda}}]{ito2020}%
  \BibitemOpen
  \bibfield  {author} {\bibinfo {author} {\bibfnamefont {S.}~\bibnamefont
  {Ito}}, \bibinfo {author} {\bibfnamefont {M.}~\bibnamefont {Arita}}, \bibinfo
  {author} {\bibfnamefont {J.}~\bibnamefont {Haruyama}}, \bibinfo {author}
  {\bibfnamefont {B.}~\bibnamefont {Feng}}, \bibinfo {author} {\bibfnamefont
  {W.-C.}\ \bibnamefont {Chen}}, \bibinfo {author} {\bibfnamefont
  {H.}~\bibnamefont {Namatame}}, \bibinfo {author} {\bibfnamefont
  {M.}~\bibnamefont {Taniguchi}}, \bibinfo {author} {\bibfnamefont {C.-M.}\
  \bibnamefont {Cheng}}, \bibinfo {author} {\bibfnamefont {G.}~\bibnamefont
  {Bian}}, \bibinfo {author} {\bibfnamefont {S.-J.}\ \bibnamefont {Tang}},
  \bibinfo {author} {\bibfnamefont {T.-C.}\ \bibnamefont {Chiang}}, \bibinfo
  {author} {\bibfnamefont {O.}~\bibnamefont {Sugino}}, \bibinfo {author}
  {\bibfnamefont {F.}~\bibnamefont {Komori}}, \ and\ \bibinfo {author}
  {\bibfnamefont {I.}~\bibnamefont {Matsuda}},\ }\bibfield  {title} {\enquote
  {\bibinfo {title} {Surface-state {Coulomb} repulsion accelerates a
  metal-insulator transition in topological semimetal nanofilms},}\ }\href@noop
  {} {\bibfield  {journal} {\bibinfo  {journal} {Sci. Adv.}\ }\textbf {\bibinfo
  {volume} {6}},\ \bibinfo {pages} {eaaz5015} (\bibinfo {year}
  {2020})}\BibitemShut {NoStop}%
\bibitem [{\citenamefont {Chang}\ \emph {et~al.}(2019)\citenamefont {Chang},
  \citenamefont {Lu}, \citenamefont {Wang}, \citenamefont {Lin}, \citenamefont
  {Miller}, \citenamefont {Chiang},\ and\ \citenamefont {Bian}}]{chang2019}%
  \BibitemOpen
  \bibfield  {author} {\bibinfo {author} {\bibfnamefont {T.-R.}\ \bibnamefont
  {Chang}}, \bibinfo {author} {\bibfnamefont {Q.}~\bibnamefont {Lu}}, \bibinfo
  {author} {\bibfnamefont {X.}~\bibnamefont {Wang}}, \bibinfo {author}
  {\bibfnamefont {H.}~\bibnamefont {Lin}}, \bibinfo {author} {\bibfnamefont
  {T.}~\bibnamefont {Miller}}, \bibinfo {author} {\bibfnamefont {T.-C.}\
  \bibnamefont {Chiang}}, \ and\ \bibinfo {author} {\bibfnamefont
  {G.}~\bibnamefont {Bian}},\ }\bibfield  {title} {\enquote {\bibinfo {title}
  {Band topology of bismuth quantum films},}\ }\href@noop {} {\bibfield
  {journal} {\bibinfo  {journal} {Crystals}\ }\textbf {\bibinfo {volume} {9}},\
  \bibinfo {pages} {510} (\bibinfo {year} {2019})}\BibitemShut {NoStop}%
\bibitem [{\citenamefont {Liu}\ \emph {et~al.}(2011)\citenamefont {Liu},
  \citenamefont {Liu}, \citenamefont {Wu}, \citenamefont {Duan}, \citenamefont
  {Liu},\ and\ \citenamefont {Wu}}]{liu2011}%
  \BibitemOpen
  \bibfield  {author} {\bibinfo {author} {\bibfnamefont {Z.}~\bibnamefont
  {Liu}}, \bibinfo {author} {\bibfnamefont {C.-X.}\ \bibnamefont {Liu}},
  \bibinfo {author} {\bibfnamefont {Y.-S.}\ \bibnamefont {Wu}}, \bibinfo
  {author} {\bibfnamefont {W.-H.}\ \bibnamefont {Duan}}, \bibinfo {author}
  {\bibfnamefont {F.}~\bibnamefont {Liu}}, \ and\ \bibinfo {author}
  {\bibfnamefont {J.}~\bibnamefont {Wu}},\ }\bibfield  {title} {\enquote
  {\bibinfo {title} {Stable nontrivial {Z}$_2$ topology in ultrathin
  {Bi}\hkl(111) films: {A} first-principles study},}\ }\href@noop {} {\bibfield
   {journal} {\bibinfo  {journal} {Phys. Rev. Lett.}\ }\textbf {\bibinfo
  {volume} {107}},\ \bibinfo {pages} {136805} (\bibinfo {year}
  {2011})}\BibitemShut {NoStop}%
\bibitem [{\citenamefont {Schindler}\ \emph {et~al.}(2018)\citenamefont
  {Schindler}, \citenamefont {Wang}, \citenamefont {Vergniory}, \citenamefont
  {Cook}, \citenamefont {Murani}, \citenamefont {Sengupta}, \citenamefont
  {Kasumov}, \citenamefont {Deblock}, \citenamefont {Jeon}, \citenamefont
  {Drozdov}, \citenamefont {Bouchiat}, \citenamefont {Guéron}, \citenamefont
  {Yazdani}, \citenamefont {Bernevig},\ and\ \citenamefont
  {Neupert}}]{schindler2018}%
  \BibitemOpen
  \bibfield  {author} {\bibinfo {author} {\bibfnamefont {F.}~\bibnamefont
  {Schindler}}, \bibinfo {author} {\bibfnamefont {Z.}~\bibnamefont {Wang}},
  \bibinfo {author} {\bibfnamefont {M.~G.}\ \bibnamefont {Vergniory}}, \bibinfo
  {author} {\bibfnamefont {A.~M.}\ \bibnamefont {Cook}}, \bibinfo {author}
  {\bibfnamefont {A.}~\bibnamefont {Murani}}, \bibinfo {author} {\bibfnamefont
  {S.}~\bibnamefont {Sengupta}}, \bibinfo {author} {\bibfnamefont {A.~Y.}\
  \bibnamefont {Kasumov}}, \bibinfo {author} {\bibfnamefont {R.}~\bibnamefont
  {Deblock}}, \bibinfo {author} {\bibfnamefont {S.}~\bibnamefont {Jeon}},
  \bibinfo {author} {\bibfnamefont {I.}~\bibnamefont {Drozdov}}, \bibinfo
  {author} {\bibfnamefont {H.}~\bibnamefont {Bouchiat}}, \bibinfo {author}
  {\bibfnamefont {S.}~\bibnamefont {Guéron}}, \bibinfo {author} {\bibfnamefont
  {A.}~\bibnamefont {Yazdani}}, \bibinfo {author} {\bibfnamefont {B.~A.}\
  \bibnamefont {Bernevig}}, \ and\ \bibinfo {author} {\bibfnamefont
  {T.}~\bibnamefont {Neupert}},\ }\bibfield  {title} {\enquote {\bibinfo
  {title} {Higher-order topology in bismuth},}\ }\href@noop {} {\bibfield
  {journal} {\bibinfo  {journal} {Nat. Phys.}\ }\textbf {\bibinfo {volume}
  {14}},\ \bibinfo {pages} {918} (\bibinfo {year} {2018})}\BibitemShut
  {NoStop}%
\bibitem [{\citenamefont {Hsu}\ \emph {et~al.}(2019)\citenamefont {Hsu},
  \citenamefont {Zhou}, \citenamefont {Chang}, \citenamefont {Ma},
  \citenamefont {Gedik}, \citenamefont {Bansil}, \citenamefont {Xu},
  \citenamefont {Lin},\ and\ \citenamefont {Fu}}]{hsu2019}%
  \BibitemOpen
  \bibfield  {author} {\bibinfo {author} {\bibfnamefont {C.-H.}\ \bibnamefont
  {Hsu}}, \bibinfo {author} {\bibfnamefont {X.}~\bibnamefont {Zhou}}, \bibinfo
  {author} {\bibfnamefont {T.-R.}\ \bibnamefont {Chang}}, \bibinfo {author}
  {\bibfnamefont {Q.}~\bibnamefont {Ma}}, \bibinfo {author} {\bibfnamefont
  {N.}~\bibnamefont {Gedik}}, \bibinfo {author} {\bibfnamefont
  {A.}~\bibnamefont {Bansil}}, \bibinfo {author} {\bibfnamefont {S.-Y.}\
  \bibnamefont {Xu}}, \bibinfo {author} {\bibfnamefont {H.}~\bibnamefont
  {Lin}}, \ and\ \bibinfo {author} {\bibfnamefont {L.}~\bibnamefont {Fu}},\
  }\bibfield  {title} {\enquote {\bibinfo {title} {Topology on a new facet of
  bismuth},}\ }\href@noop {} {\bibfield  {journal} {\bibinfo  {journal} {Proc.
  Natl. Acad. Sci. U.S.A.}\ }\textbf {\bibinfo {volume} {116}},\ \bibinfo
  {pages} {13255} (\bibinfo {year} {2019})}\BibitemShut {NoStop}%
\bibitem [{\citenamefont {Tang}\ \emph {et~al.}(2019)\citenamefont {Tang},
  \citenamefont {Po}, \citenamefont {Vishwanath},\ and\ \citenamefont
  {Wan}}]{tang2019}%
  \BibitemOpen
  \bibfield  {author} {\bibinfo {author} {\bibfnamefont {F.}~\bibnamefont
  {Tang}}, \bibinfo {author} {\bibfnamefont {H.~C.}\ \bibnamefont {Po}},
  \bibinfo {author} {\bibfnamefont {A.}~\bibnamefont {Vishwanath}}, \ and\
  \bibinfo {author} {\bibfnamefont {X.}~\bibnamefont {Wan}},\ }\bibfield
  {title} {\enquote {\bibinfo {title} {Comprehensive search for topological
  materials using symmetry indicators},}\ }\href@noop {} {\bibfield  {journal}
  {\bibinfo  {journal} {Nature}\ }\textbf {\bibinfo {volume} {566}},\ \bibinfo
  {pages} {486} (\bibinfo {year} {2019})}\BibitemShut {NoStop}%
\bibitem [{\citenamefont {Zhang}\ \emph {et~al.}(2019)\citenamefont {Zhang},
  \citenamefont {Jiang}, \citenamefont {Song}, \citenamefont {Huang},
  \citenamefont {He}, \citenamefont {Fang}, \citenamefont {Weng},\ and\
  \citenamefont {Fang}}]{zhang2019}%
  \BibitemOpen
  \bibfield  {author} {\bibinfo {author} {\bibfnamefont {T.}~\bibnamefont
  {Zhang}}, \bibinfo {author} {\bibfnamefont {Y.}~\bibnamefont {Jiang}},
  \bibinfo {author} {\bibfnamefont {Z.}~\bibnamefont {Song}}, \bibinfo {author}
  {\bibfnamefont {H.}~\bibnamefont {Huang}}, \bibinfo {author} {\bibfnamefont
  {Y.}~\bibnamefont {He}}, \bibinfo {author} {\bibfnamefont {Z.}~\bibnamefont
  {Fang}}, \bibinfo {author} {\bibfnamefont {H.}~\bibnamefont {Weng}}, \ and\
  \bibinfo {author} {\bibfnamefont {C.}~\bibnamefont {Fang}},\ }\bibfield
  {title} {\enquote {\bibinfo {title} {Catalogue of topological electronic
  materials},}\ }\href@noop {} {\bibfield  {journal} {\bibinfo  {journal}
  {Nature}\ }\textbf {\bibinfo {volume} {566}},\ \bibinfo {pages} {475}
  (\bibinfo {year} {2019})}\BibitemShut {NoStop}%
\bibitem [{\citenamefont {Abdelbarey}\ \emph {et~al.}(2020)\citenamefont
  {Abdelbarey}, \citenamefont {Koch}, \citenamefont {Mamiyev}, \citenamefont
  {Tegenkamp},\ and\ \citenamefont {Pfnür}}]{abdelbarey2020}%
  \BibitemOpen
  \bibfield  {author} {\bibinfo {author} {\bibfnamefont {D.}~\bibnamefont
  {Abdelbarey}}, \bibinfo {author} {\bibfnamefont {J.}~\bibnamefont {Koch}},
  \bibinfo {author} {\bibfnamefont {Z.}~\bibnamefont {Mamiyev}}, \bibinfo
  {author} {\bibfnamefont {C.}~\bibnamefont {Tegenkamp}}, \ and\ \bibinfo
  {author} {\bibfnamefont {H.}~\bibnamefont {Pfnür}},\ }\bibfield  {title}
  {\enquote {\bibinfo {title} {Thickness-dependent electronic transport through
  epitaxial nontrivial {Bi} quantum films},}\ }\href@noop {} {\bibfield
  {journal} {\bibinfo  {journal} {Phys. Rev. B}\ }\textbf {\bibinfo {volume}
  {102}},\ \bibinfo {pages} {115409} (\bibinfo {year} {2020})}\BibitemShut
  {NoStop}%
\bibitem [{\citenamefont {Jin}\ \emph {et~al.}(2020)\citenamefont {Jin},
  \citenamefont {Yeom},\ and\ \citenamefont {Liu}}]{jin2020}%
  \BibitemOpen
  \bibfield  {author} {\bibinfo {author} {\bibfnamefont {K.-H.}\ \bibnamefont
  {Jin}}, \bibinfo {author} {\bibfnamefont {H.~W.}\ \bibnamefont {Yeom}}, \
  and\ \bibinfo {author} {\bibfnamefont {F.}~\bibnamefont {Liu}},\ }\bibfield
  {title} {\enquote {\bibinfo {title} {Doping-induced topological phase
  transition in {Bi}: The role of quantum electronic stress},}\ }\href@noop {}
  {\bibfield  {journal} {\bibinfo  {journal} {Phys. Rev. B}\ }\textbf {\bibinfo
  {volume} {101}},\ \bibinfo {pages} {035111} (\bibinfo {year}
  {2020})}\BibitemShut {NoStop}%
\bibitem [{\citenamefont {Moore}\ and\ \citenamefont
  {Orenstein}(2010)}]{moore2010}%
  \BibitemOpen
  \bibfield  {author} {\bibinfo {author} {\bibfnamefont {J.~E.}\ \bibnamefont
  {Moore}}\ and\ \bibinfo {author} {\bibfnamefont {J.}~\bibnamefont
  {Orenstein}},\ }\bibfield  {title} {\enquote {\bibinfo {title}
  {Confinement-induced {Berry} phase and helicity-dependent photocurrents},}\
  }\href@noop {} {\bibfield  {journal} {\bibinfo  {journal} {Phys. Rev. Lett.}\
  }\textbf {\bibinfo {volume} {105}},\ \bibinfo {pages} {026805} (\bibinfo
  {year} {2010})}\BibitemShut {NoStop}%
\bibitem [{\citenamefont {Sawahata}\ \emph {et~al.}(2018)\citenamefont
  {Sawahata}, \citenamefont {Yamaguchi}, \citenamefont {Kotaka},\ and\
  \citenamefont {Ishii}}]{sawahata2018}%
  \BibitemOpen
  \bibfield  {author} {\bibinfo {author} {\bibfnamefont {H.}~\bibnamefont
  {Sawahata}}, \bibinfo {author} {\bibfnamefont {N.}~\bibnamefont {Yamaguchi}},
  \bibinfo {author} {\bibfnamefont {H.}~\bibnamefont {Kotaka}}, \ and\ \bibinfo
  {author} {\bibfnamefont {F.}~\bibnamefont {Ishii}},\ }\bibfield  {title}
  {\enquote {\bibinfo {title} {First-principles study of electric-field-induced
  topological phase transition in one-bilayer {Bi}\hkl(111)},}\ }\href@noop {}
  {\bibfield  {journal} {\bibinfo  {journal} {Jpn. J. Appl. Phys.}\ }\textbf
  {\bibinfo {volume} {57}},\ \bibinfo {pages} {030309} (\bibinfo {year}
  {2018})}\BibitemShut {NoStop}%
\bibitem [{\citenamefont {Schiferl}\ and\ \citenamefont
  {Barrett}(1969)}]{schiferl1969}%
  \BibitemOpen
  \bibfield  {author} {\bibinfo {author} {\bibfnamefont {D.}~\bibnamefont
  {Schiferl}}\ and\ \bibinfo {author} {\bibfnamefont {C.~S.}\ \bibnamefont
  {Barrett}},\ }\bibfield  {title} {\enquote {\bibinfo {title} {The crystal
  structure of arsenic at $4.2$, $78$ and $299$\si{\degree\kelvin}},}\
  }\href@noop {} {\bibfield  {journal} {\bibinfo  {journal} {J. Appl. Cryst.}\
  }\textbf {\bibinfo {volume} {2}},\ \bibinfo {pages} {30} (\bibinfo {year}
  {1969})}\BibitemShut {NoStop}%
\bibitem [{\citenamefont {Timrov}\ \emph {et~al.}(2012)\citenamefont {Timrov},
  \citenamefont {Kampfrath}, \citenamefont {Faure}, \citenamefont {Vast},
  \citenamefont {Ast}, \citenamefont {Frischkorn}, \citenamefont {Wolf},
  \citenamefont {Gava},\ and\ \citenamefont {Perfetti}}]{timrov2012}%
  \BibitemOpen
  \bibfield  {author} {\bibinfo {author} {\bibfnamefont {I.}~\bibnamefont
  {Timrov}}, \bibinfo {author} {\bibfnamefont {T.}~\bibnamefont {Kampfrath}},
  \bibinfo {author} {\bibfnamefont {J.}~\bibnamefont {Faure}}, \bibinfo
  {author} {\bibfnamefont {N.}~\bibnamefont {Vast}}, \bibinfo {author}
  {\bibfnamefont {C.~R.}\ \bibnamefont {Ast}}, \bibinfo {author} {\bibfnamefont
  {C.}~\bibnamefont {Frischkorn}}, \bibinfo {author} {\bibfnamefont
  {M.}~\bibnamefont {Wolf}}, \bibinfo {author} {\bibfnamefont {P.}~\bibnamefont
  {Gava}}, \ and\ \bibinfo {author} {\bibfnamefont {L.}~\bibnamefont
  {Perfetti}},\ }\bibfield  {title} {\enquote {\bibinfo {title} {Thermalization
  of photoexcited carriers in bismuth investigated by time-resolved terahertz
  spectroscopy and ab initio calculations},}\ }\href@noop {} {\bibfield
  {journal} {\bibinfo  {journal} {Phys. Rev. B}\ }\textbf {\bibinfo {volume}
  {85}},\ \bibinfo {pages} {155139} (\bibinfo {year} {2012})}\BibitemShut
  {NoStop}%
\bibitem [{\citenamefont {Giannozzi}\ \emph {et~al.}(2009)\citenamefont
  {Giannozzi}, \citenamefont {Baroni}, \citenamefont {Bonini}, \citenamefont
  {Calandra}, \citenamefont {Car}, \citenamefont {Cavazzoni}, \citenamefont
  {Ceresoli}, \citenamefont {Chiarotti}, \citenamefont {Cococcioni},
  \citenamefont {Dabo}, \citenamefont {Dal~Corso}, \citenamefont
  {de~Gironcoli}, \citenamefont {Fabris}, \citenamefont {Fratesi},
  \citenamefont {Gebauer}, \citenamefont {Gerstmann}, \citenamefont
  {Gougoussis}, \citenamefont {Kokalj}, \citenamefont {Lazzeri}, \citenamefont
  {Martin-Samos}, \citenamefont {Marzari}, \citenamefont {Mauri}, \citenamefont
  {Mazzarello}, \citenamefont {Paolini}, \citenamefont {Pasquarello},
  \citenamefont {Paulatto}, \citenamefont {Sbraccia}, \citenamefont {Scandolo},
  \citenamefont {Sclauzero}, \citenamefont {Seitsonen}, \citenamefont
  {Smogunov}, \citenamefont {Umari},\ and\ \citenamefont
  {Wentzcovitch}}]{giannozzi2009}%
  \BibitemOpen
  \bibfield  {author} {\bibinfo {author} {\bibfnamefont {P.}~\bibnamefont
  {Giannozzi}}, \bibinfo {author} {\bibfnamefont {S.}~\bibnamefont {Baroni}},
  \bibinfo {author} {\bibfnamefont {N.}~\bibnamefont {Bonini}}, \bibinfo
  {author} {\bibfnamefont {M.}~\bibnamefont {Calandra}}, \bibinfo {author}
  {\bibfnamefont {R.}~\bibnamefont {Car}}, \bibinfo {author} {\bibfnamefont
  {C.}~\bibnamefont {Cavazzoni}}, \bibinfo {author} {\bibfnamefont
  {D.}~\bibnamefont {Ceresoli}}, \bibinfo {author} {\bibfnamefont {G.~L.}\
  \bibnamefont {Chiarotti}}, \bibinfo {author} {\bibfnamefont {M.}~\bibnamefont
  {Cococcioni}}, \bibinfo {author} {\bibfnamefont {I.}~\bibnamefont {Dabo}},
  \bibinfo {author} {\bibfnamefont {A.}~\bibnamefont {Dal~Corso}}, \bibinfo
  {author} {\bibfnamefont {S.}~\bibnamefont {de~Gironcoli}}, \bibinfo {author}
  {\bibfnamefont {S.}~\bibnamefont {Fabris}}, \bibinfo {author} {\bibfnamefont
  {G.}~\bibnamefont {Fratesi}}, \bibinfo {author} {\bibfnamefont
  {R.}~\bibnamefont {Gebauer}}, \bibinfo {author} {\bibfnamefont
  {U.}~\bibnamefont {Gerstmann}}, \bibinfo {author} {\bibfnamefont
  {C.}~\bibnamefont {Gougoussis}}, \bibinfo {author} {\bibfnamefont
  {A.}~\bibnamefont {Kokalj}}, \bibinfo {author} {\bibfnamefont
  {M.}~\bibnamefont {Lazzeri}}, \bibinfo {author} {\bibfnamefont
  {L.}~\bibnamefont {Martin-Samos}}, \bibinfo {author} {\bibfnamefont
  {N.}~\bibnamefont {Marzari}}, \bibinfo {author} {\bibfnamefont
  {F.}~\bibnamefont {Mauri}}, \bibinfo {author} {\bibfnamefont
  {R.}~\bibnamefont {Mazzarello}}, \bibinfo {author} {\bibfnamefont
  {S.}~\bibnamefont {Paolini}}, \bibinfo {author} {\bibfnamefont
  {A.}~\bibnamefont {Pasquarello}}, \bibinfo {author} {\bibfnamefont
  {L.}~\bibnamefont {Paulatto}}, \bibinfo {author} {\bibfnamefont
  {C.}~\bibnamefont {Sbraccia}}, \bibinfo {author} {\bibfnamefont
  {S.}~\bibnamefont {Scandolo}}, \bibinfo {author} {\bibfnamefont
  {G.}~\bibnamefont {Sclauzero}}, \bibinfo {author} {\bibfnamefont {A.~P.}\
  \bibnamefont {Seitsonen}}, \bibinfo {author} {\bibfnamefont {A.}~\bibnamefont
  {Smogunov}}, \bibinfo {author} {\bibfnamefont {P.}~\bibnamefont {Umari}}, \
  and\ \bibinfo {author} {\bibfnamefont {R.~M.}\ \bibnamefont {Wentzcovitch}},\
  }\bibfield  {title} {\enquote {\bibinfo {title} {\textsc{Quantum Espresso}: A
  modular and open-source software project for quantum simulations of
  materials},}\ }\href@noop {} {\bibfield  {journal} {\bibinfo  {journal} {J.
  Phys.: Condens. Matter}\ }\textbf {\bibinfo {volume} {21}},\ \bibinfo {pages}
  {395502} (\bibinfo {year} {2009})}\BibitemShut {NoStop}%
\bibitem [{\citenamefont {Giannozzi}\ \emph {et~al.}(2017)\citenamefont
  {Giannozzi}, \citenamefont {Andreussi}, \citenamefont {Brumme}, \citenamefont
  {Bunau}, \citenamefont {Buongiorno~Nardelli}, \citenamefont {Calandra},
  \citenamefont {Car}, \citenamefont {Cavazzoni}, \citenamefont {Ceresoli},
  \citenamefont {Cococcioni}, \citenamefont {Colonna}, \citenamefont
  {Carnimeo}, \citenamefont {Dal~Corso}, \citenamefont {de~Gironcoli},
  \citenamefont {Delugas}, \citenamefont {DiStasio~Jr.}, \citenamefont
  {Ferretti}, \citenamefont {Floris}, \citenamefont {Fratesi}, \citenamefont
  {Fugallo}, \citenamefont {Gebauer}, \citenamefont {Gerstmann}, \citenamefont
  {Giustino}, \citenamefont {Gorni}, \citenamefont {Jia}, \citenamefont
  {Kawamura}, \citenamefont {Ko}, \citenamefont {Kokalj}, \citenamefont
  {Kü{\c{c}}ükbenli}, \citenamefont {Lazzeri}, \citenamefont {Marsili},
  \citenamefont {Marzari}, \citenamefont {Mauri}, \citenamefont {Nguyen},
  \citenamefont {Nguyen}, \citenamefont {{Otero-de-la-Roza}}, \citenamefont
  {Paulatto}, \citenamefont {Ponc{\'{e}}}, \citenamefont {Rocca}, \citenamefont
  {Sabatini}, \citenamefont {Santra}, \citenamefont {Schlipf}, \citenamefont
  {Seitsonen}, \citenamefont {Smogunov}, \citenamefont {Timrov}, \citenamefont
  {Thonhauser}, \citenamefont {Umari}, \citenamefont {Vast}, \citenamefont
  {Wu},\ and\ \citenamefont {Baroni}}]{giannozzi2017}%
  \BibitemOpen
  \bibfield  {author} {\bibinfo {author} {\bibfnamefont {P.}~\bibnamefont
  {Giannozzi}}, \bibinfo {author} {\bibfnamefont {O.}~\bibnamefont
  {Andreussi}}, \bibinfo {author} {\bibfnamefont {T.}~\bibnamefont {Brumme}},
  \bibinfo {author} {\bibfnamefont {O.}~\bibnamefont {Bunau}}, \bibinfo
  {author} {\bibfnamefont {M.}~\bibnamefont {Buongiorno~Nardelli}}, \bibinfo
  {author} {\bibfnamefont {M.}~\bibnamefont {Calandra}}, \bibinfo {author}
  {\bibfnamefont {R.}~\bibnamefont {Car}}, \bibinfo {author} {\bibfnamefont
  {C.}~\bibnamefont {Cavazzoni}}, \bibinfo {author} {\bibfnamefont
  {D.}~\bibnamefont {Ceresoli}}, \bibinfo {author} {\bibfnamefont
  {M.}~\bibnamefont {Cococcioni}}, \bibinfo {author} {\bibfnamefont
  {N.}~\bibnamefont {Colonna}}, \bibinfo {author} {\bibfnamefont
  {I.}~\bibnamefont {Carnimeo}}, \bibinfo {author} {\bibfnamefont
  {A.}~\bibnamefont {Dal~Corso}}, \bibinfo {author} {\bibfnamefont
  {S.}~\bibnamefont {de~Gironcoli}}, \bibinfo {author} {\bibfnamefont
  {P.}~\bibnamefont {Delugas}}, \bibinfo {author} {\bibfnamefont {R.~A.}\
  \bibnamefont {DiStasio~Jr.}}, \bibinfo {author} {\bibfnamefont
  {A.}~\bibnamefont {Ferretti}}, \bibinfo {author} {\bibfnamefont
  {A.}~\bibnamefont {Floris}}, \bibinfo {author} {\bibfnamefont
  {G.}~\bibnamefont {Fratesi}}, \bibinfo {author} {\bibfnamefont
  {G.}~\bibnamefont {Fugallo}}, \bibinfo {author} {\bibfnamefont
  {R.}~\bibnamefont {Gebauer}}, \bibinfo {author} {\bibfnamefont
  {U.}~\bibnamefont {Gerstmann}}, \bibinfo {author} {\bibfnamefont
  {F.}~\bibnamefont {Giustino}}, \bibinfo {author} {\bibfnamefont
  {T.}~\bibnamefont {Gorni}}, \bibinfo {author} {\bibfnamefont
  {J.}~\bibnamefont {Jia}}, \bibinfo {author} {\bibfnamefont {M.}~\bibnamefont
  {Kawamura}}, \bibinfo {author} {\bibfnamefont {H.-Y.}\ \bibnamefont {Ko}},
  \bibinfo {author} {\bibfnamefont {A.}~\bibnamefont {Kokalj}}, \bibinfo
  {author} {\bibfnamefont {E.}~\bibnamefont {Kü{\c{c}}ükbenli}}, \bibinfo
  {author} {\bibfnamefont {M.}~\bibnamefont {Lazzeri}}, \bibinfo {author}
  {\bibfnamefont {M.}~\bibnamefont {Marsili}}, \bibinfo {author} {\bibfnamefont
  {N.}~\bibnamefont {Marzari}}, \bibinfo {author} {\bibfnamefont
  {F.}~\bibnamefont {Mauri}}, \bibinfo {author} {\bibfnamefont {N.~L.}\
  \bibnamefont {Nguyen}}, \bibinfo {author} {\bibfnamefont {H.-V.}\
  \bibnamefont {Nguyen}}, \bibinfo {author} {\bibfnamefont {A.}~\bibnamefont
  {{Otero-de-la-Roza}}}, \bibinfo {author} {\bibfnamefont {L.}~\bibnamefont
  {Paulatto}}, \bibinfo {author} {\bibfnamefont {S.}~\bibnamefont
  {Ponc{\'{e}}}}, \bibinfo {author} {\bibfnamefont {D.}~\bibnamefont {Rocca}},
  \bibinfo {author} {\bibfnamefont {R.}~\bibnamefont {Sabatini}}, \bibinfo
  {author} {\bibfnamefont {B.}~\bibnamefont {Santra}}, \bibinfo {author}
  {\bibfnamefont {M.}~\bibnamefont {Schlipf}}, \bibinfo {author} {\bibfnamefont
  {A.~P.}\ \bibnamefont {Seitsonen}}, \bibinfo {author} {\bibfnamefont
  {A.}~\bibnamefont {Smogunov}}, \bibinfo {author} {\bibfnamefont
  {I.}~\bibnamefont {Timrov}}, \bibinfo {author} {\bibfnamefont
  {T.}~\bibnamefont {Thonhauser}}, \bibinfo {author} {\bibfnamefont
  {P.}~\bibnamefont {Umari}}, \bibinfo {author} {\bibfnamefont
  {N.}~\bibnamefont {Vast}}, \bibinfo {author} {\bibfnamefont {X.}~\bibnamefont
  {Wu}}, \ and\ \bibinfo {author} {\bibfnamefont {S.}~\bibnamefont {Baroni}},\
  }\bibfield  {title} {\enquote {\bibinfo {title} {Advanced capabilities for
  materials modelling with \textsc{Quantum Espresso}},}\ }\href@noop {}
  {\bibfield  {journal} {\bibinfo  {journal} {J. Phys.: Condens. Matter}\
  }\textbf {\bibinfo {volume} {29}},\ \bibinfo {pages} {465901} (\bibinfo
  {year} {2017})}\BibitemShut {NoStop}%
\bibitem [{\citenamefont {Hamann}(2013)}]{hamann2013}%
  \BibitemOpen
  \bibfield  {author} {\bibinfo {author} {\bibfnamefont {D.~R.}\ \bibnamefont
  {Hamann}},\ }\bibfield  {title} {\enquote {\bibinfo {title} {Optimized
  norm-conserving {Vanderbilt} pseudopotentials},}\ }\href@noop {} {\bibfield
  {journal} {\bibinfo  {journal} {Phys. Rev. B}\ }\textbf {\bibinfo {volume}
  {88}},\ \bibinfo {pages} {085117} (\bibinfo {year} {2013})}\BibitemShut
  {NoStop}%
\bibitem [{\citenamefont {Schlipf}\ and\ \citenamefont
  {Gygi}(2015)}]{schlipf2015}%
  \BibitemOpen
  \bibfield  {author} {\bibinfo {author} {\bibfnamefont {M.}~\bibnamefont
  {Schlipf}}\ and\ \bibinfo {author} {\bibfnamefont {F.}~\bibnamefont {Gygi}},\
  }\bibfield  {title} {\enquote {\bibinfo {title} {Optimization algorithm for
  the generation of {ONCV} pseudopotentials},}\ }\href@noop {} {\bibfield
  {journal} {\bibinfo  {journal} {Comput. Phys. Commun.}\ }\textbf {\bibinfo
  {volume} {196}},\ \bibinfo {pages} {36} (\bibinfo {year} {2015})}\BibitemShut
  {NoStop}%
\bibitem [{\citenamefont {Scherpelz}\ \emph {et~al.}(2016)\citenamefont
  {Scherpelz}, \citenamefont {Govoni}, \citenamefont {Hamada},\ and\
  \citenamefont {Galli}}]{scherpelz2016}%
  \BibitemOpen
  \bibfield  {author} {\bibinfo {author} {\bibfnamefont {P.}~\bibnamefont
  {Scherpelz}}, \bibinfo {author} {\bibfnamefont {M.}~\bibnamefont {Govoni}},
  \bibinfo {author} {\bibfnamefont {I.}~\bibnamefont {Hamada}}, \ and\ \bibinfo
  {author} {\bibfnamefont {G.}~\bibnamefont {Galli}},\ }\bibfield  {title}
  {\enquote {\bibinfo {title} {Implementation and validation of fully
  relativistic {GW} calculations: {Spin}-orbit coupling in molecules,
  nanocrystals, and solids},}\ }\href@noop {} {\bibfield  {journal} {\bibinfo
  {journal} {J. Chem. Theory Comput.}\ }\textbf {\bibinfo {volume} {12}},\
  \bibinfo {pages} {3523} (\bibinfo {year} {2016})}\BibitemShut {NoStop}%
\bibitem [{\citenamefont {Perdew}\ \emph {et~al.}(1996)\citenamefont {Perdew},
  \citenamefont {Burke},\ and\ \citenamefont {Ernzerhof}}]{perdew1996}%
  \BibitemOpen
  \bibfield  {author} {\bibinfo {author} {\bibfnamefont {J.~P.}\ \bibnamefont
  {Perdew}}, \bibinfo {author} {\bibfnamefont {K.}~\bibnamefont {Burke}}, \
  and\ \bibinfo {author} {\bibfnamefont {M.}~\bibnamefont {Ernzerhof}},\
  }\bibfield  {title} {\enquote {\bibinfo {title} {Generalized gradient
  approximation made simple},}\ }\href@noop {} {\bibfield  {journal} {\bibinfo
  {journal} {Phys. Rev. Lett.}\ }\textbf {\bibinfo {volume} {77}},\ \bibinfo
  {pages} {3865} (\bibinfo {year} {1996})}\BibitemShut {NoStop}%
\bibitem [{\citenamefont {Marini}\ \emph {et~al.}(2009)\citenamefont {Marini},
  \citenamefont {Hogan}, \citenamefont {Grüning},\ and\ \citenamefont
  {Varsano}}]{marini2009}%
  \BibitemOpen
  \bibfield  {author} {\bibinfo {author} {\bibfnamefont {A.}~\bibnamefont
  {Marini}}, \bibinfo {author} {\bibfnamefont {C.}~\bibnamefont {Hogan}},
  \bibinfo {author} {\bibfnamefont {M.}~\bibnamefont {Grüning}}, \ and\
  \bibinfo {author} {\bibfnamefont {D.}~\bibnamefont {Varsano}},\ }\bibfield
  {title} {\enquote {\bibinfo {title} {\textsc{Yambo}: An ab initio tool for
  excited state calculations},}\ }\href@noop {} {\bibfield  {journal} {\bibinfo
   {journal} {Comput. Phys. Commun.}\ }\textbf {\bibinfo {volume} {180}},\
  \bibinfo {pages} {1392} (\bibinfo {year} {2009})}\BibitemShut {NoStop}%
\bibitem [{\citenamefont {Sangalli}\ \emph {et~al.}(2019)\citenamefont
  {Sangalli}, \citenamefont {Ferretti}, \citenamefont {Miranda}, \citenamefont
  {Attaccalite}, \citenamefont {Marri}, \citenamefont {Cannuccia},
  \citenamefont {Melo}, \citenamefont {Marsili}, \citenamefont {Paleari},
  \citenamefont {Marrazzo}, \citenamefont {Prandini}, \citenamefont
  {Bonf{\`{a}}}, \citenamefont {Atambo}, \citenamefont {Affinito},
  \citenamefont {Palummo}, \citenamefont {Molina-S{\'{a}}nchez}, \citenamefont
  {Hogan}, \citenamefont {Grüning}, \citenamefont {Varsano},\ and\
  \citenamefont {Marini}}]{sangalli2019}%
  \BibitemOpen
  \bibfield  {author} {\bibinfo {author} {\bibfnamefont {D.}~\bibnamefont
  {Sangalli}}, \bibinfo {author} {\bibfnamefont {A.}~\bibnamefont {Ferretti}},
  \bibinfo {author} {\bibfnamefont {H.}~\bibnamefont {Miranda}}, \bibinfo
  {author} {\bibfnamefont {C.}~\bibnamefont {Attaccalite}}, \bibinfo {author}
  {\bibfnamefont {I.}~\bibnamefont {Marri}}, \bibinfo {author} {\bibfnamefont
  {E.}~\bibnamefont {Cannuccia}}, \bibinfo {author} {\bibfnamefont
  {P.}~\bibnamefont {Melo}}, \bibinfo {author} {\bibfnamefont {M.}~\bibnamefont
  {Marsili}}, \bibinfo {author} {\bibfnamefont {F.}~\bibnamefont {Paleari}},
  \bibinfo {author} {\bibfnamefont {A.}~\bibnamefont {Marrazzo}}, \bibinfo
  {author} {\bibfnamefont {G.}~\bibnamefont {Prandini}}, \bibinfo {author}
  {\bibfnamefont {P.}~\bibnamefont {Bonf{\`{a}}}}, \bibinfo {author}
  {\bibfnamefont {M.~O.}\ \bibnamefont {Atambo}}, \bibinfo {author}
  {\bibfnamefont {F.}~\bibnamefont {Affinito}}, \bibinfo {author}
  {\bibfnamefont {M.}~\bibnamefont {Palummo}}, \bibinfo {author} {\bibfnamefont
  {A.}~\bibnamefont {Molina-S{\'{a}}nchez}}, \bibinfo {author} {\bibfnamefont
  {C.}~\bibnamefont {Hogan}}, \bibinfo {author} {\bibfnamefont
  {M.}~\bibnamefont {Grüning}}, \bibinfo {author} {\bibfnamefont
  {D.}~\bibnamefont {Varsano}}, \ and\ \bibinfo {author} {\bibfnamefont
  {A.}~\bibnamefont {Marini}},\ }\bibfield  {title} {\enquote {\bibinfo {title}
  {Many-body perturbation theory calculations using the \textsc{Yambo} code},}\
  }\href@noop {} {\bibfield  {journal} {\bibinfo  {journal} {J. Phys.: Condens.
  Matter}\ }\textbf {\bibinfo {volume} {31}},\ \bibinfo {pages} {325902}
  (\bibinfo {year} {2019})}\BibitemShut {NoStop}%
\bibitem [{\citenamefont {Sakuma}\ \emph {et~al.}(2011)\citenamefont {Sakuma},
  \citenamefont {Friedrich}, \citenamefont {Miyake}, \citenamefont {Blügel},\
  and\ \citenamefont {Aryasetiawan}}]{sakuma2011}%
  \BibitemOpen
  \bibfield  {author} {\bibinfo {author} {\bibfnamefont {R.}~\bibnamefont
  {Sakuma}}, \bibinfo {author} {\bibfnamefont {C.}~\bibnamefont {Friedrich}},
  \bibinfo {author} {\bibfnamefont {T.}~\bibnamefont {Miyake}}, \bibinfo
  {author} {\bibfnamefont {S.}~\bibnamefont {Blügel}}, \ and\ \bibinfo
  {author} {\bibfnamefont {F.}~\bibnamefont {Aryasetiawan}},\ }\bibfield
  {title} {\enquote {\bibinfo {title} {{GW} calculations including spin-orbit
  coupling: {Application} to {Hg} chalcogenides},}\ }\href@noop {} {\bibfield
  {journal} {\bibinfo  {journal} {Phys. Rev. B}\ }\textbf {\bibinfo {volume}
  {84}},\ \bibinfo {pages} {085144} (\bibinfo {year} {2011})}\BibitemShut
  {NoStop}%
\bibitem [{\citenamefont {Bruneval}\ and\ \citenamefont
  {Gonze}(2008)}]{bruneval2008}%
  \BibitemOpen
  \bibfield  {author} {\bibinfo {author} {\bibfnamefont {F.}~\bibnamefont
  {Bruneval}}\ and\ \bibinfo {author} {\bibfnamefont {X.}~\bibnamefont
  {Gonze}},\ }\bibfield  {title} {\enquote {\bibinfo {title} {Accurate {GW}
  self-energies in a plane-wave basis using only a few empty states: Towards
  large systems},}\ }\href@noop {} {\bibfield  {journal} {\bibinfo  {journal}
  {Phys. Rev. B}\ }\textbf {\bibinfo {volume} {78}},\ \bibinfo {pages} {085125}
  (\bibinfo {year} {2008})}\BibitemShut {NoStop}%
\bibitem [{\citenamefont {Madsen}\ and\ \citenamefont
  {Singh}(2006)}]{madsen2006}%
  \BibitemOpen
  \bibfield  {author} {\bibinfo {author} {\bibfnamefont {G.~K.~H.}\
  \bibnamefont {Madsen}}\ and\ \bibinfo {author} {\bibfnamefont {D.~J.}\
  \bibnamefont {Singh}},\ }\bibfield  {title} {\enquote {\bibinfo {title}
  {\textsc{BoltzTraP}. {A} code for calculating band-structure dependent
  quantities},}\ }\href@noop {} {\bibfield  {journal} {\bibinfo  {journal}
  {Comput. Phys. Commun.}\ }\textbf {\bibinfo {volume} {175}},\ \bibinfo
  {pages} {67} (\bibinfo {year} {2006})}\BibitemShut {NoStop}%
\bibitem [{\citenamefont {Madsen}\ \emph {et~al.}(2018)\citenamefont {Madsen},
  \citenamefont {Carrete},\ and\ \citenamefont {Verstraete}}]{madsen2018}%
  \BibitemOpen
  \bibfield  {author} {\bibinfo {author} {\bibfnamefont {G.~K.~H.}\
  \bibnamefont {Madsen}}, \bibinfo {author} {\bibfnamefont {J.}~\bibnamefont
  {Carrete}}, \ and\ \bibinfo {author} {\bibfnamefont {M.~J.}\ \bibnamefont
  {Verstraete}},\ }\bibfield  {title} {\enquote {\bibinfo {title}
  {\textsc{BoltzTraP2}, a program for interpolating band structures and
  calculating semi-classical transport coefficients},}\ }\href@noop {}
  {\bibfield  {journal} {\bibinfo  {journal} {Comput. Phys. Commun.}\ }\textbf
  {\bibinfo {volume} {231}},\ \bibinfo {pages} {140} (\bibinfo {year}
  {2018})}\BibitemShut {NoStop}%
\bibitem [{Note2()}]{Note2}%
  \BibitemOpen
  \bibinfo {note} {We note that the cost of exactly fitting the energies at the
  grid points is that minor Gibbs oscillations occur in the band energies
  between grid points.}\BibitemShut {Stop}%
\bibitem [{\citenamefont {Hasan}\ and\ \citenamefont {Kane}(2010)}]{hasan2010}%
  \BibitemOpen
  \bibfield  {author} {\bibinfo {author} {\bibfnamefont {M.~Z.}\ \bibnamefont
  {Hasan}}\ and\ \bibinfo {author} {\bibfnamefont {C.~L.}\ \bibnamefont
  {Kane}},\ }\bibfield  {title} {\enquote {\bibinfo {title} {Colloquium:
  {Topological} insulators},}\ }\href@noop {} {\bibfield  {journal} {\bibinfo
  {journal} {Rev. Mod. Phys.}\ }\textbf {\bibinfo {volume} {82}},\ \bibinfo
  {pages} {3045} (\bibinfo {year} {2010})}\BibitemShut {NoStop}%
\bibitem [{Note3()}]{Note3}%
  \BibitemOpen
  \bibinfo {note} {Further details will be given below but it is already
  apparent how band inversions play a role here since they turn valence bands
  into conduction bands and vice versa.}\BibitemShut {Stop}%
\bibitem [{sup({\natexlab{a}})}]{supplemental-bands}%
  \BibitemOpen
  \href@noop {} {} ({\natexlab{a}}),\ \bibinfo {note} {see Supplemental
  Material at URL2 for the bulk band structure as a function of
  $a_r$.}\BibitemShut {Stop}%
\bibitem [{Note4()}]{Note4}%
  \BibitemOpen
  \bibinfo {note} {$\nu _1$: L$_1$, X$_1$, X$_3$, T$_1$; $\nu _2$: L$_2$,
  X$_1$, X$_2$, T$_1$; $\nu _3$: L$_3$, X$_2$, X$_3$, T$_1$; Kramer's
  degenerate states are considered only once.}\BibitemShut {Stop}%
\bibitem [{Note5()}]{Note5}%
  \BibitemOpen
  \bibinfo {note} {$a_h$ defines the width of the cell in-plane with the
  bilayers. $c_h$ is the out-of-plane coordinate and thus controls the distance
  between the bilayers. Typically, thin Bi films grow so that the bilayers lie
  flat on the substrate. For a conversion between the rhombohedral and
  hexagonal cell see for example Ref.~\cite {hofmann2006}.}\BibitemShut {Stop}%
\bibitem [{\citenamefont {Nagao}\ \emph {et~al.}(2004)\citenamefont {Nagao},
  \citenamefont {Sadowski}, \citenamefont {Saito}, \citenamefont {Yaginuma},
  \citenamefont {Fujikawa}, \citenamefont {Kogure}, \citenamefont {Ohno},
  \citenamefont {Hasegawa}, \citenamefont {Hasegawa},\ and\ \citenamefont
  {Sakurai}}]{nagao2004}%
  \BibitemOpen
  \bibfield  {author} {\bibinfo {author} {\bibfnamefont {T.}~\bibnamefont
  {Nagao}}, \bibinfo {author} {\bibfnamefont {J.~T.}\ \bibnamefont {Sadowski}},
  \bibinfo {author} {\bibfnamefont {M.}~\bibnamefont {Saito}}, \bibinfo
  {author} {\bibfnamefont {S.}~\bibnamefont {Yaginuma}}, \bibinfo {author}
  {\bibfnamefont {Y.}~\bibnamefont {Fujikawa}}, \bibinfo {author}
  {\bibfnamefont {T.}~\bibnamefont {Kogure}}, \bibinfo {author} {\bibfnamefont
  {T.}~\bibnamefont {Ohno}}, \bibinfo {author} {\bibfnamefont {Y.}~\bibnamefont
  {Hasegawa}}, \bibinfo {author} {\bibfnamefont {S.}~\bibnamefont {Hasegawa}},
  \ and\ \bibinfo {author} {\bibfnamefont {T.}~\bibnamefont {Sakurai}},\
  }\bibfield  {title} {\enquote {\bibinfo {title} {Nanofilm allotrope and phase
  transformation of ultrathin {Bi} film on {Si}(111)-7$\times$7},}\ }\href@noop
  {} {\bibfield  {journal} {\bibinfo  {journal} {Phys. Rev. Lett.}\ }\textbf
  {\bibinfo {volume} {93}},\ \bibinfo {pages} {105501} (\bibinfo {year}
  {2004})}\BibitemShut {NoStop}%
\bibitem [{sup({\natexlab{b}})}]{supplemental-main}%
  \BibitemOpen
  \href@noop {} {} ({\natexlab{b}}),\ \bibinfo {note} {see Supplemental
  Material at URL1 for the LDA results and additional details.}\BibitemShut
  {Stop}%
\bibitem [{\citenamefont {Aguado-Puente}\ \emph {et~al.}(2020)\citenamefont
  {Aguado-Puente}, \citenamefont {Fahy},\ and\ \citenamefont
  {Grüning}}]{aguadopuente2020}%
  \BibitemOpen
  \bibfield  {author} {\bibinfo {author} {\bibfnamefont {P.}~\bibnamefont
  {Aguado-Puente}}, \bibinfo {author} {\bibfnamefont {S.}~\bibnamefont {Fahy}},
  \ and\ \bibinfo {author} {\bibfnamefont {M.}~\bibnamefont {Grüning}},\
  }\bibfield  {title} {\enquote {\bibinfo {title} {{GW} study of
  pressure-induced topological insulator transition in group-{IV}
  tellurides},}\ }\href@noop {} {\bibfield  {journal} {\bibinfo  {journal}
  {Phys. Rev. Res.}\ }\textbf {\bibinfo {volume} {2}},\ \bibinfo {pages}
  {043105} (\bibinfo {year} {2020})}\BibitemShut {NoStop}%
\bibitem [{\citenamefont {Golin}(1968)}]{golin1968}%
  \BibitemOpen
  \bibfield  {author} {\bibinfo {author} {\bibfnamefont {S.}~\bibnamefont
  {Golin}},\ }\bibfield  {title} {\enquote {\bibinfo {title} {Band structure of
  bismuth: Pseudopotential approach},}\ }\href@noop {} {\bibfield  {journal}
  {\bibinfo  {journal} {Phys. Rev.}\ }\textbf {\bibinfo {volume} {166}},\
  \bibinfo {pages} {643} (\bibinfo {year} {1968})}\BibitemShut {NoStop}%
\bibitem [{\citenamefont {Hofmann}(2006)}]{hofmann2006}%
  \BibitemOpen
  \bibfield  {author} {\bibinfo {author} {\bibfnamefont {P.}~\bibnamefont
  {Hofmann}},\ }\bibfield  {title} {\enquote {\bibinfo {title} {The surfaces of
  bismuth: Structural and electronic properties},}\ }\href@noop {} {\bibfield
  {journal} {\bibinfo  {journal} {Prog. Surf. Sci.}\ }\textbf {\bibinfo
  {volume} {81}},\ \bibinfo {pages} {191} (\bibinfo {year} {2006})}\BibitemShut
  {NoStop}%
\end{thebibliography}%

\end{document}


\title{\textit{Supplemental Material:} Effect of strain and many-body corrections on the band inversions and topology of bismuth}

\author{Christian \surname{König}}
\affiliation{Tyndall National Institute, University College Cork, Lee Maltings, Cork T12 R5CP, Ireland}
\author{James C. \surname{Greer}}
\affiliation{Nottingham Ningbo New Material Institute and Department of Electrical and Electronic Engineering, University of Nottingham Ningbo China, 199 Taikang East Road, Ningbo, 315100, China}
\author{Stephen \surname{Fahy}}
\affiliation{Tyndall National Institute, University College Cork, Lee Maltings, Cork T12 R5CP, Ireland}
\affiliation{Department of Physics, University College Cork, College Road, Cork T12 K8AF, Ireland}

\date{\today}

\begin{abstract}
We here present some more data which may be of interest to the reader.
In particular we plot the fitted values for the internal displacement and show how it affects the main features of the electronic structure.
The LDA results are also shown for comparison with the GGA.
Furthermore, the remaining G$_0$W$_0$ results for the overlap are included for completeness.
Finally, the band structures of the expanded cells in Fig.~2 of the main text are provided in a separate file.
\end{abstract}

\maketitle

\begin{figure*} 
  \center
  \includegraphics{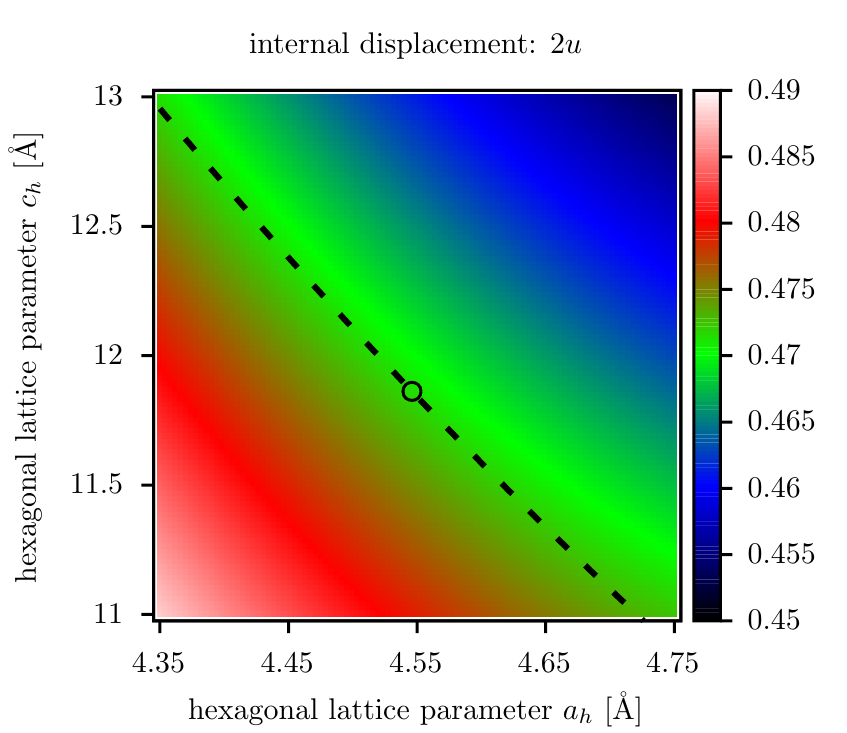}
  \caption{
  Plot of the internal displacement as a function of the hexagonal lattice parameters $a_h$ and $c_h$.
  We show the value of $2u$, i.e.~the smallest distance between the two basis atoms along the trigonal axis.
  For $2u=0.5$ there is no displacement of the atom from the middle of the cell.
  As we can see, under the assumption of constant volume, keeping $u$ fixed is a good assumption.
}
\end{figure*}

\begin{figure*} 
\center
\includegraphics{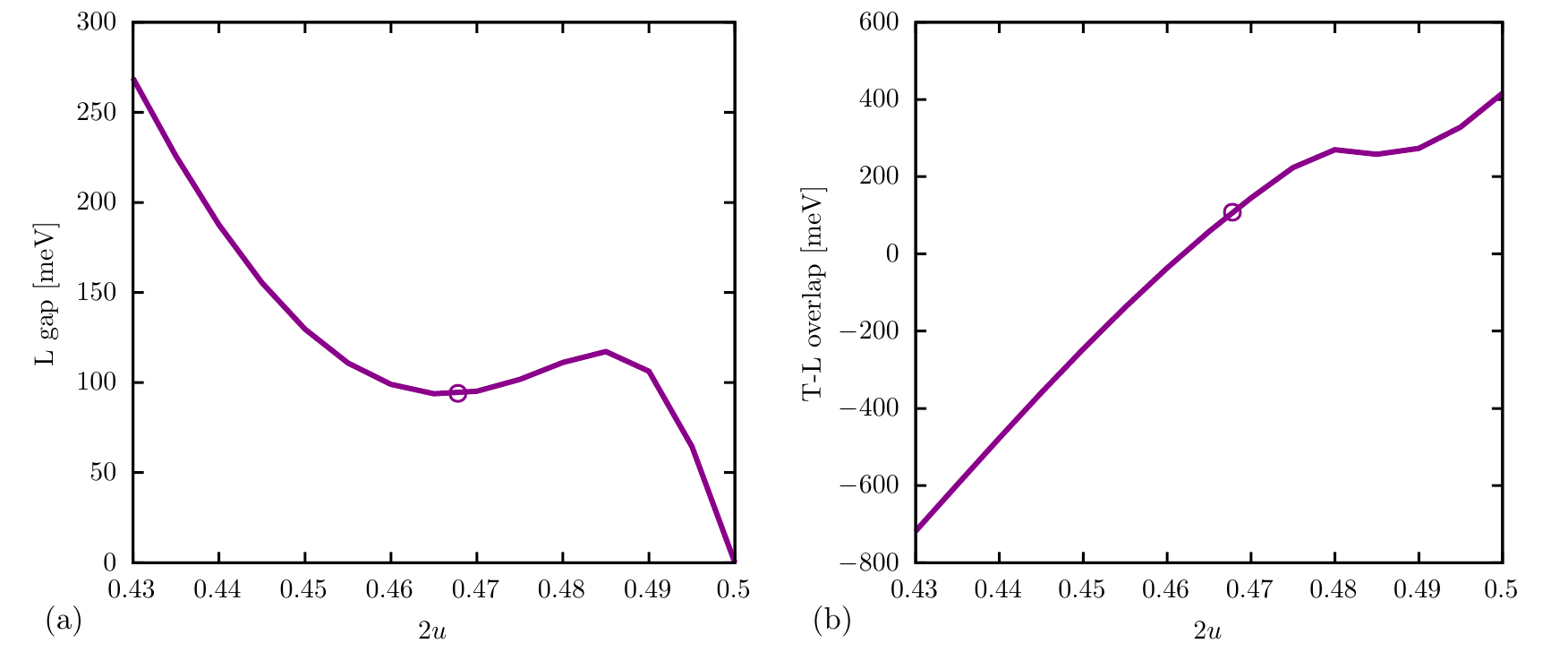}
\caption{
  The effect of the internal displacement on (a) the L gap and (b) the indirect T-L overlap is shown.
  The results for the experimental displacement of $2u=0.46778$ are highlighted with a circle.
  The remaining two cell parameters $a_r$ and $\alpha$ were not changed and we used their literature values for all calculations.
  We find that the optimized $2u=0.47131$ is close to the experimental value and the minimum L gap and that a small variation, e.g.~due to imperfect relaxation,
  is likely to increase the gap but otherwise does not have dramatic effects.
  The overlap on the other hand changes more dramatically.
  Approaching $2u=0.5$ brings the structure closer to the cubic limit (although still $\alpha \neq \SI{60}{\degree}$) and the L gap vanishes completely.
}
\end{figure*}

\begin{figure*} 
\center
\includegraphics{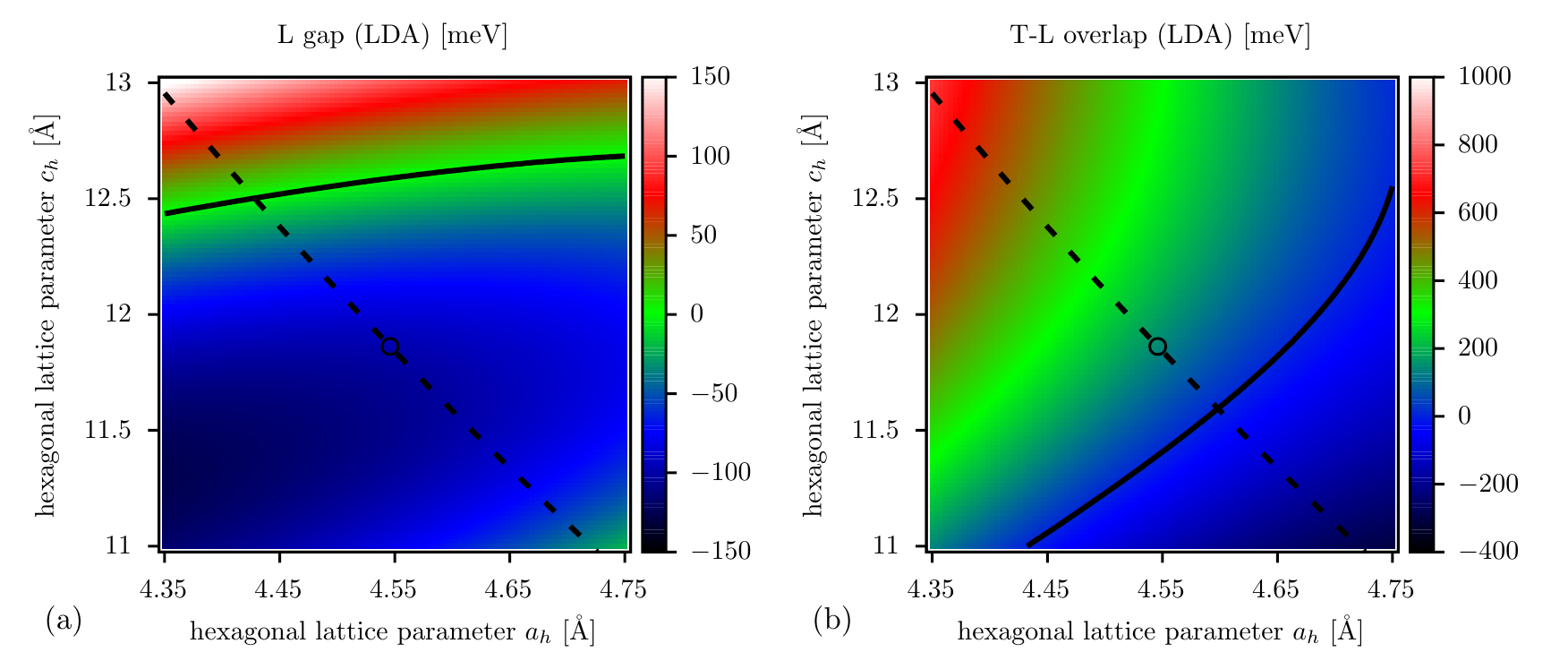}
\caption{
  LDA results for the L gap and T-L overlap. There is no significant qualitative difference to the GGA.
}
\end{figure*}

\begin{figure*} 
  \center
  \includegraphics{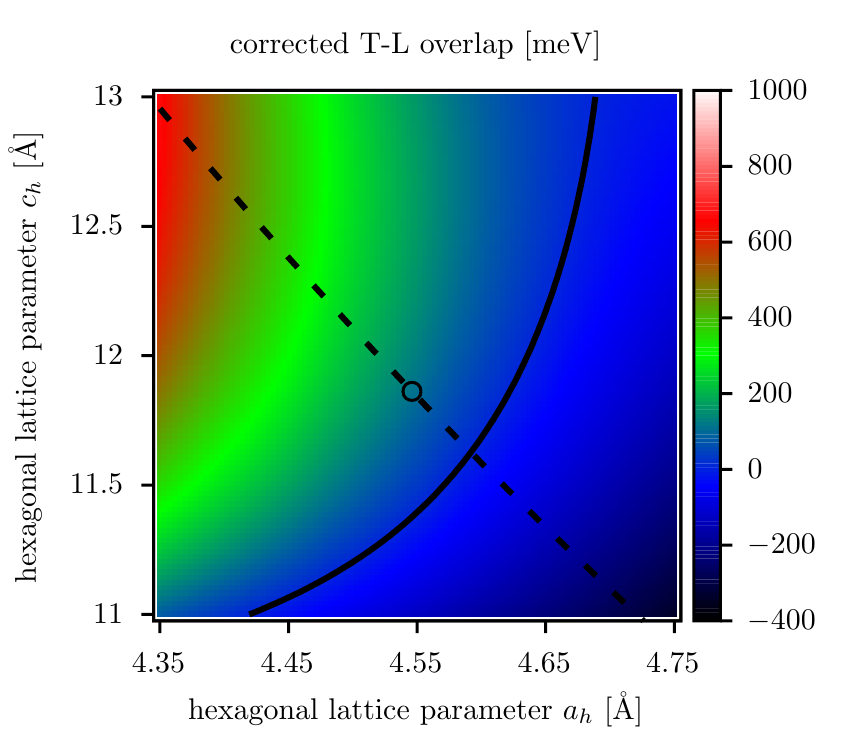}
  \caption{
  Plot of the indirect overlap with G$_0$W$_0$ corrections.
}
\end{figure*}